\documentclass[iop]{emulateapj}

\slugcomment{Accepted for publication in the Astrophysical Journal}

\shorttitle{Extremely-bright submm galaxies}
\shortauthors{Tamura et al.}

\begin{document}


\title{Extremely-bright submillimeter galaxies beyond the Lupus-I star-forming region}

\author{
	Y. Tamura\altaffilmark{1}, 
	R. Kawabe\altaffilmark{2,3,4}, 
	Y. Shimajiri\altaffilmark{5,6}, 
	T. Tsukagoshi\altaffilmark{7}, 
	Y. Nakajima\altaffilmark{8}, 
	Y. Oasa\altaffilmark{9}, 
	D.~J. Wilner\altaffilmark{10}, 
	C.~J. Chandler\altaffilmark{11}, 
	K. Saigo\altaffilmark{12},
	K. Tomida\altaffilmark{13,14},
	M.~S. Yun\altaffilmark{15}, 
	A. Taniguchi\altaffilmark{1},
	K. Kohno\altaffilmark{1,16},
	B. Hatsukade\altaffilmark{12}, 
	I. Aretxaga\altaffilmark{17},
	J.~E. Austermann\altaffilmark{18},
	R. Dickman\altaffilmark{11}, 
	H. Ezawa\altaffilmark{12}, 
	W.~M. Goss\altaffilmark{11}, 
	M. Hayashi\altaffilmark{2,19},
	D.~H. Hughes\altaffilmark{17}, 
	M. Hiramatsu\altaffilmark{12},
	S. Inutsuka\altaffilmark{20},
	R. Ogasawara\altaffilmark{12},
	N. Ohashi\altaffilmark{21},
	T. Oshima\altaffilmark{6},
	K.~S. Scott\altaffilmark{22},
	and
	G.~W. Wilson\altaffilmark{15}
}

\altaffiltext{1}{Institute of Astronomy, The University of Tokyo, Osawa, Mitaka, Tokyo 181-0015, Japan;  \email{ytamura@ioa.s.u-tokyo.ac.jp}}
\altaffiltext{2}{National Astronomical Observatory of Japan, Osawa, Mitaka, Tokyo 181-8588, Japan}
\altaffiltext{3}{Department of Astronomy, School of Science, The Graduate University for Advanced Studies (SOKENDAI), Osawa, Mitaka, Tokyo 181-8588, Japan} 
\altaffiltext{4}{Joint ALMA Observatory, Alonso de Cordova 3107 Vitacura, Santiago 763 0355, Chile}
\altaffiltext{5}{Laboratoire AIM, CEA/DSM-CNRS-Universit\'{e} Paris Diderot, IRFU/Service d'Astrophysique, CEA Saclay, F-91191 Gif-sur-Yvette, France}
\altaffiltext{6}{Nobeyama Radio Observatory, National Astronomical Observatory of Japan, Nobeyama, Minamimaki, Minamisaku, Nagano 384-1305, Japan}
\altaffiltext{7}{Institute of Astrophysics and Planetary Sciences, Ibaraki University, 2-1-1 Bunkyo, Mito, Ibaraki 310-8512, Japan}
\altaffiltext{8}{Hitotsubashi University, 2-1 Naka, Kunitachi, Tokyo 186-8601, Japan}
\altaffiltext{9}{Faculty of Education, Saitama University, 255 Shimo-Okubo, Sakura, Saitama, Saitama 388-8570, Japan}
\altaffiltext{10}{Harvard-Smithsonian Center for Astrophysics, 60 Garden Street, Cambridge, MA 02138}
\altaffiltext{11}{National Radio Astronomy Observatory, P.O. Box 0, Socorro, NM 87801, USA}
\altaffiltext{12}{Chile Observatory, National Astronomical Observatory of Japan, Osawa, Mitaka, Tokyo 181-8588, Japan}
\altaffiltext{13}{Department of Astrophysical Sciences, Princeton University, Princeton, NJ 08544}
\altaffiltext{14}{Department of Physics, The University of Tokyo, Hongo, Bunkyo, Tokyo 113-0033, Japan}
\altaffiltext{15}{Department of Astronomy, University of Massachusetts, 710 North Pleasant Street, Amherst, MA 01003}
\altaffiltext{16}{Research Center for the Early Universe, The University of Tokyo, Hongo, Bunkyo, Tokyo 113-0033, Japan}
\altaffiltext{17}{Instituto Nacional de Astrofisica, Optica y Electronica, Aptdo.\ Postal 51 y 216, 72000 Puebla, Mexico}
\altaffiltext{18}{Center for Astrophysics and Space Astronomy, University of Colorado, Boulder, CO 80309}
\altaffiltext{19}{School of Mathematical and Physical Science, The Graduate University for Advanced Studies (SOKENDAI), Hayama, Kanagawa 240-0193, Japan}
\altaffiltext{20}{Department of Physics, Nagoya University, Nagoya, Aichi 464-8602, Japan}
\altaffiltext{21}{Subaru Telescope, National Astronomical Observatory of Japan, 650 North A'ohoku Place, Hilo, HI 96720}
\altaffiltext{22}{North American ALMA Science Center, National Radio Astronomy Observatory, 520 Edgemont Road, Charlottesville, Virginia 22903}


\begin{abstract}
We report detections of two candidate distant submillimeter galaxies (SMGs), MM~J154506.4$-$344318 and MM~J154132.7$-$350320, which are discovered in the AzTEC/ASTE 1.1~mm survey toward the Lupus-I star-forming region.  The two objects have 1.1~mm flux densities of 43.9 and 27.1~mJy, and have {\it Herschel}/SPIRE counterparts as well.  The Submillimeter Array counterpart to the former SMG is identified at 890~$\micron$ and 1.3~mm.  Photometric redshift estimates using all available data from the mid-infrared to the radio suggest that the redshifts of the two SMGs are $z_{\rm photo} \simeq 4$--5 and 3, respectively.  Near-infrared objects are found very close to the SMGs and they are consistent with low-$z$ ellipticals, suggesting that the high apparent luminosities can be attributed to gravitational magnification.  The cumulative number counts at $S_{\rm 1.1mm} \ge 25$~mJy, combined with other two 1.1-mm brightest sources, are $0.70 ^{+0.56}_{-0.34}$~deg$^{-2}$, which is consistent with a model prediction that accounts for flux magnification due to strong  gravitational lensing.  Unexpectedly, a $z > 3$ SMG and a Galactic dense starless core (e.g., a first hydrostatic core) could be similar in the mid-infrared to millimeter spectral energy distributions and spatial structures at least at $\gtrsim 1''$.  This indicates that it is necessary to distinguish the two possibilities by means of broad band photometry from the optical to centimeter and spectroscopy to determine the redshift, when a compact object is identified toward Galactic star-forming regions.
\end{abstract}



\section{Introduction}

In recent years, remarkable progress has been made in finding the \emph{brightest} ($S_{\rm 850 \mu m} \gtrsim 70$ mJy or $S_{\rm 1.1mm} \gtrsim 30$ mJy) population of submillimeter (submm) galaxies \citep[SMGs, e.g.,][]{Blain02,Casey14}  via square-degree scale surveys using far-infrared (FIR) to millimeter (mm) single dish telescopes in space and on the ground.  Thanks to their apparent high luminosity, usually with the aid of gravitational magnification, the brightest SMGs offer a unique opportunity to investigate physical properties of the ISM \citep[e.g.,][]{Harris10, Ivison10, Cox11, Danielson11, Danielson13, Scott11, Valtchanov11, Combes12, Decarli12, Iono12, Lupu12, Bothwell13, Omont13}, resolved star-forming activity \citep[e.g.,][]{Swinbank10, Negrello10, Fu12, Fu13} and gas dynamics \citep[e.g.,][]{Riechers11, Rawle13, Messias14} at the peak of the star-formation history of galaxies. 

Furthermore, the brightest SMGs located in the \emph{high-redshift tail} of the SMG redshift distribution \citep[median $z \approx 2$--3, e.g.,][]{Chapman05, Yun12, Swinbank14} at $z = 3$--6 provide a severe challenge to theories of galaxy formation and evolution \citep{Baugh05, Granato04, Granato06}.  The surface density of the brightest SMGs holds integrated information on when galaxies undergo intense starburst and how frequently strong gravitational lensing occurs, on which galaxy formation models depend.  But, the source counts of brightest SMGs in the high-$z$ tail are highly uncertain because of poor statistics, and it is indeed hard to find them; only a single SMG with $S_{\rm 1.1mm} \gtrsim 30$~mJy is expected within $\sim 1$--10~deg$^{2}$ from limited knowledge of current studies \citep{Scott12, Vieira10}.  Large-area surveys using \emph{Herschel}/SPIRE (250, 350, and 500 $\mu$m) have recently revealed strongly lensed galaxies bright at 250--500~$\micron$ \citep[e.g.,][]{Negrello10, Gonzalez-Nuevo12}.  These surveys bring about a great change in our understanding of the statistical properties of the SMG population.   The SPIRE bands, however, preferentially select SMGs at modest redshifts ($z \lesssim 3$), and the study of SMGs in the high-$z$ tail ($z = 3$--6) is still far from being complete.

The mm to long-submm wavelength cameras such as SCUBA, SCUBA-2 \citep[850 $\micron$,][]{Holland99, Holland13}, MAMBO-2 \citep[1.2 mm,][]{Kreysa98}, AzTEC \citep[1.1 mm,][]{Wilson08a}, and Laboca \citep[870 $\micron$,][]{Siringo09} are complementary to those FIR to short-submm surveys;  For example, observations at longer wavelengths with the MAMBO and AzTEC cameras and the South Pole Telescope  \citep[SPT,][]{Carlstrom11} may exploit a uniform selection function in redshift space, allowing the efficient detection of the brightest SMGs out to $z > 3$ \citep[e.g.,][]{Lestrade09, Lestrade10, Ikarashi11, Vieira13, Weiss13}.   We have used the AzTEC camera to carry out large-area surveys toward Galactic star-forming regions, which cover $\sim 30$~deg$^{2}$ of the sky in total with $1\sigma$ sensitivities of 5--30~mJy~beam$^{-1}$.  From mm/submm number counts \citep{Negrello10, Vieira10, Scott12, Takekoshi13}, several detections of ultra-bright ($S_{\rm 1.1mm} \gtrsim 30$~mJy) extragalactic sources at cosmological distances are expected within the survey area.
This is complementary to earlier attempts to search for extremely-bright SMGs and to constrain the brightest end of  extragalactic number counts by exploiting submm maps from Galactic surveys \citep[e.g.,][]{Barnard04}.

\begin{figure}[t]
	\begin{center}
		\includegraphics[scale=0.68]{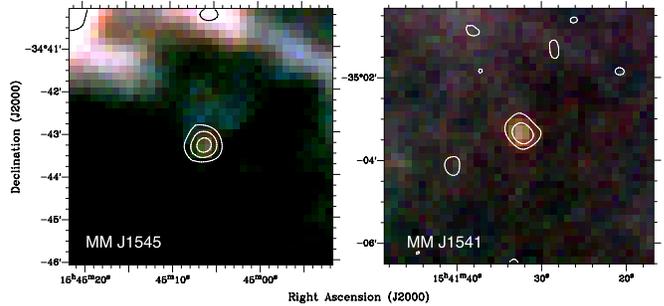}
		\caption{
			The $6' \times 6'$ far-infrared to mm images of MM-J1545 (left) and MM-J1541 (right).  
			The contours show the AzTEC 1.1-mm image, starting from $2\sigma$ 
			with a separation of $2\sigma$.  The background images show 
			the \emph{Herschel}/SPIRE 250, 350, 500-$\micron$ pseudo-color images. 
		}\label{fig:azteconspire}
	\end{center}
\end{figure}

In this paper, we report the AzTEC detections and multi-wavelength analyses of two 1.1-mm bright sources, MM~J154506.4$-$344318 and MM~J154132.7$-$350320 (hereafter MM-J1545 and MM-J1541, respectively; see Figure~\ref{fig:azteconspire}).  These sources are found toward the Lupus-I star-forming region, a local ($z = 0$) molecular cloud, and indeed the close proximity of MM-J1545 to the molecular cloud misled us to classify it as a starless core when it was initially identified with the AzTEC 1.1-mm camera.  Multi-wavelength data collected by subsequent follow-up observations, however, strongly suggest that they are galaxies at cosmological distances as presented in this paper, illustrating the importance of multi-wavelength imaging and spectroscopy of such objects.  

The paper is organized as follows. Section~\ref{sect:obsdata} describes observations and archival data obtained toward MM-J1545 and MM-J1541.  In Section~\ref{sect:results}, we present the results from the observations and photometric redshift estimates.  Section~\ref{sect:discussions} discusses the gravitational lensing effect on both objects, the brightest end of the 1.1~mm number counts, and the FIR-to-mm colors of the mm-selected sources.  In Section~\ref{sect:discussions}, we also explore a possibility that MM-J1545 would be  a Galactic starless core.  Finally, Section~\ref{sect:conclusions} summarizes our conclusions. 

Throughout this paper, we assume a concordance cosmology with $\Omega _{\rm m} = 0.3$, $\Omega _{\Lambda} = 0.7$, and $H_0 = 70$ km~s$^{-1}$~Mpc$^{-1}$.


\begin{deluxetable*}{llcccc}
\tablewidth{13.5cm}
\tablecaption{Multiwavelength counterparts to MM-J1545 and MM-J1541. \label{tab1_Multiwavelength}}
\tablehead{
	\colhead{Instrument} & \colhead{Band} &\multicolumn{2}{c}{Flux density} & \colhead{Unit} & \colhead{Ref} \\
	 &  & \colhead{MM-J1545} & \colhead{MM-J1541} &  & 
}
\startdata
VLA  &  20 cm  & ... & $1.3 \pm 0.5$ & mJy  & 1 \\
VLA  &  6 cm  & $66 \pm 5$ & ... & $\mu$Jy  & 2 \\
ATCA & 7 mm & $210 \pm 35$ & ... & $\mu$Jy & 2 \\
NMA  &  2.7 mm  & $< 5.9$ & ... & mJy  & 2 \\
SMA  &  1.3 mm  &  $20.8 \pm 1.9$\tablenotemark{b} & ... & mJy & 2 \\
AzTEC  & 1.1 mm  &  $43.9 \pm 5.6$\tablenotemark{a} & $27.1 \pm 5.0$\tablenotemark{a} & mJy & 2, 3 \\
SMA  &  890 $\micron$  &  $69.7 \pm 12.1$\tablenotemark{b} & ... & mJy & 2  \\
SPIRE  &  500 $\micron$  &  $134.9 \pm 11.9$\tablenotemark{g} & $150.5 \pm 12.8$\tablenotemark{g} & mJy & 4  \\
  &  350 $\micron$  &  $109.4 \pm 11.4$\tablenotemark{g} & $169.5 \pm 11.8$\tablenotemark{g} & mJy & 4  \\
  &  250 $\micron$  &  $40.9 \pm 12.7$\tablenotemark{g} & $144.0 \pm 18.0$\tablenotemark{g} & mJy & 4  \\
MIPS  & 24 $\micron$ &  $< 0.3$ & $1.51 \pm 0.31$\tablenotemark{c}  & mJy & 5 \\
IRAC   & 5.8 $\micron$  &  $< 4.5$ & ... & $\mu$Jy & 5 \\
 & 3.6 $\micron$  & $< 0.7$ & ... & $\mu$Jy & 5 \\
MOIRCS & 2.15 $\micron$ & $< 2.7$\tablenotemark{d} & ... & $\mu$Jy & 2 \\
WFCAM & 1.64 $\micron$ & $< 7.8$\tablenotemark{d} & ... & $\mu$Jy & 2  \\
 & 1.26 $\micron$  & $< 10.2$\tablenotemark{d} & ... & $\mu$Jy & 2 \\
2MASS & $K_\mathrm{s}$ & \multicolumn{2}{c}{$< 122$\tablenotemark{e}} & $\mu$Jy & 5 \\
 & $H$ & \multicolumn{2}{c}{$< 95$\tablenotemark{e}} & $\mu$Jy & 5 \\
 & $J$ & \multicolumn{2}{c}{$< 60$\tablenotemark{e}}& $\mu$Jy & 5 \\
DSS & $I$ & \multicolumn{2}{c}{$< 93$\tablenotemark{f}} & $\mu$Jy & 5 \\
 & $R$  & \multicolumn{2}{c}{$< 14$\tablenotemark{f}} & $\mu$Jy & 5 \\
 & $B$  & \multicolumn{2}{c}{$< 2.5$\tablenotemark{f}} & $\mu$Jy & 5 
\enddata
		\tablenotetext{a}{The flux density not corrected for the flux bias due to confusion noises.  The flux may be deboosted by only $\lesssim 10\%$ 
if the number counts are as flat as $dN/dS \propto S^{-2.5}$  at $> 10$ mJy.} 
		\tablenotetext{b}{Flux density estimated from the visibility fitting.}
		\tablenotetext{c}{The flux density retrieved from the catalog of the C2D survey \citep{Evans03, Evans09} Data Release 4, 
				in which the object is identified asSSTc2d J154132.7$-$350320}
		\tablenotetext{d}{The $2\sigma$ limiting flux with a $3\farcs 0$ aperture.}
		\tablenotetext{e}{The limit may be inaccurate due to blending of the near-IR object (see \S~\ref{sect:individual} and Table \ref{tab2_Lens}).}
		\tablenotetext{f}{$3\sigma$ limiting flux of the DSS/SERC survey.}
\tablenotetext{g}{The flux density measured with the \emph{Herschel} Interactive Processing Environment \citep[HIPE,][]{Ott10} command \texttt{SourceExtractorSussextractor}.
  Source confusion is not accounted for in the error.}
		\tablerefs{
(1) The NRAO VLA Sky Survey (NVSS), (2) This work, (3) Kawabe et al., in preparation; (4) \textit{Herschel} Science Archive (HSA); (5) NASA/IPAC Infrared Science Archive (IRSA).
		}
\end{deluxetable*}%


\begin{deluxetable}{llccc}
\tablewidth{8.5cm}
\tablecaption{Photometry of optical/near-infrared objects J1545B and J1541B. \label{tab2_Lens}} 
\tablehead{
	\colhead{Instrument} & \colhead{Band} &\multicolumn{2}{c}{Flux density} & \colhead{Unit} \\
&  & \colhead{J1545B} & \colhead{J1541B} &  
}
\startdata
IRAC & 5.8 $\micron$  &  $59.0 \pm 39.2$\tablenotemark{a} & ... & $\mu$Jy  \\
& 3.6 $\micron$  & $49.8 \pm 5.2$\tablenotemark{a} & ... & $\mu$Jy  \\
WISE  & 4.6 $\micron$ & $< 100$\tablenotemark{e} & $253 \pm 19$\tablenotemark{b} & $\mu$Jy \\
  & 3.2 $\micron$ & $< 90$\tablenotemark{e} & $260 \pm 12$\tablenotemark{b} & $\mu$Jy \\
MOIRCS & 2.15 $\micron$ & $25.0 \pm 2.1$ & ... & $\mu$Jy \\
WFCAM & 1.64 $\micron$ & $27.5 \pm 10.1$ & ... & $\mu$Jy  \\
 & 1.26 $\micron$  & $17.2 \pm 5.9$ & ... & $\mu$Jy \\
2MASS & $K_\mathrm{s}$ & ... & $231 \pm 28$\tablenotemark{c} & $\mu$Jy \\
 & $H$ & ... & $237 \pm 28$\tablenotemark{c} & $\mu$Jy \\
 & $J$ & ... & $286 \pm 34$\tablenotemark{c} & $\mu$Jy \\
DSS & $I$ & $< 93$ & $261 \pm 44$\tablenotemark{d} & $\mu$Jy \\
 & $R$  & $< 14$  & $275 \pm 33$\tablenotemark{d} & $\mu$Jy  \\
 & $B$  & $< 2.5$  & $182 \pm 22$\tablenotemark{d} & $\mu$Jy  \\
\enddata
		\tablenotetext{a}{From the catalog of the C2D survey \citep{Evans03, Evans09} Data Release 4, 
				in which the object is identified as SSTc2d J154506.4$-$344318.}
		\tablenotetext{b}{From the WISE All-Sky Source Catalog.}
		\tablenotetext{c}{From the 2MASS All-Sky Point Source Catalog, in which the object is identified as 2MASS J15413256$-$3503233.}
		\tablenotetext{d}{From the USNO-B1.0 Catalog \citep{Monet03}.  
			The uncertainty includes the statistical and systematic photometric errors.}
		\tablenotetext{e}{The limit is uncertain due to source confusion from an adjacent bright source.}
\end{deluxetable}%


\begin{figure*}[htbp]
	\begin{center}
		\includegraphics[scale=0.72,angle=-90]{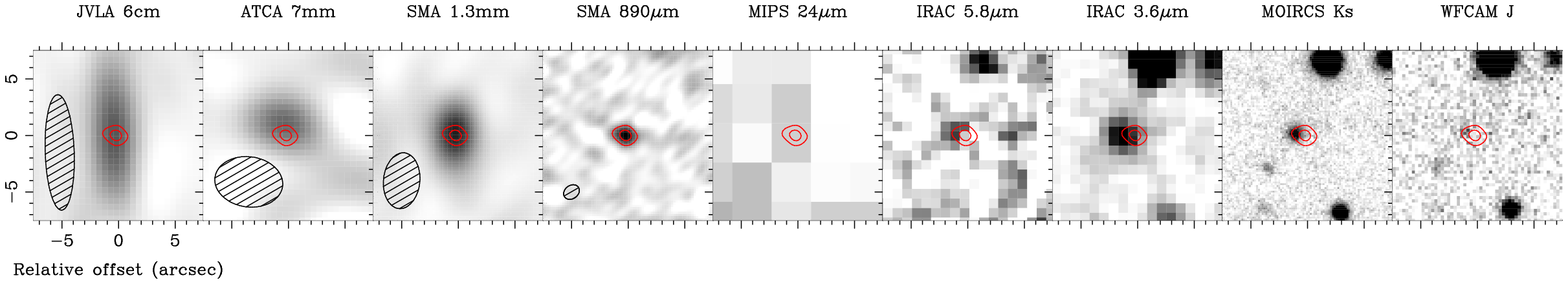}\\
		\vspace{5mm}
		\includegraphics[scale=0.72,angle=-90]{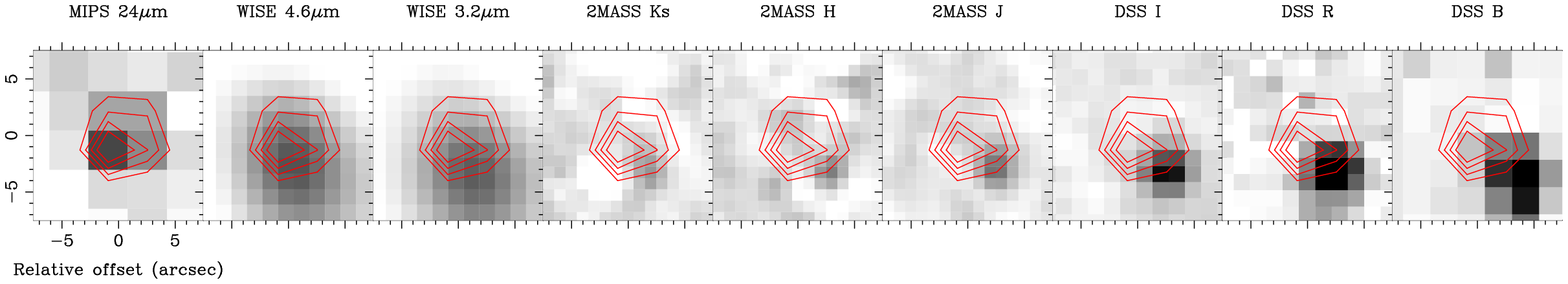}
		\caption{
			(top) The $15'' \times 15''$ postage stamp images of MM-J1545, 
			centered at the SMA 890-$\micron$ position.  From left to right 
			we show images at radio to near-infrared wavelengths.  
			The contours represent the SMA 890-$\micron$ image, 
			which are drawn at 4$\sigma$ and 8$\sigma$ ($1\sigma = 3.0$~mJy~beam$^{-1}$).  
			For the interferometer images, we also show the beam sizes with hatched ellipses.  
			An object seen in near-infrared images, referred to as J1545B,  
			is closely associated with the SMA source, but is offsets by $0\farcs 9$.  
			(bottom) The $15'' \times 15''$ postage stamp images of MM-J1541, 
			centered at the MIPS 24-$\micron$ position.  The images at mid-infrared to 
			optical wavelengths are shown.  The overlaid contours show the 24-$\micron$ image, 
			which clearly shows the systematic offset ($3''$) from a near-infrared 
			and optical source, J1541B.
		}\label{fig:stamp}
	\end{center}
\end{figure*}


\section{Observations and data}\label{sect:obsdata}

\subsection{AzTEC Observations and Archival Data}

The 1.1-mm data were taken with the AzTEC bolometer camera \citep{Wilson08a} installed on the Atacama Submillimeter Experiment 10~m telescope \citep[ASTE,][]{Ezawa04, Ezawa08} located at Atacama (altitude of 4860~m) in the Chilean Andes, during August 2007 to December 2008.  The weather conditions during the runs were excellent;  The typical 220-GHz zenith opacities were in the range of 0.01--0.08.  AzTEC/ASTE provides an angular resolution of $28''$ in full-width at half maximum (FWHM).  The complete description of the Lupus-I starless core survey will be reported elsewhere (Tsukagoshi et al., in preparation). 

The reduction procedure is described by \citet{Scott08} and \citet{Downes12}.  To extract point-like sources, the time-stream data were intensively cleaned using a principal component analysis (PCA) algorithm, and then mapped.  The PCA cleaning works as a high-pass filter in the spatial frequency domain, and thus emission extended significantly compared with the beam solid angle is fully filtered out. The FWHM of the point response function is $37''$ after an optimal filter is applied to surpress high spatial frequency noises.  The pointing was checked every 1~hr using nearby radio quasars, resulting in the astrometric accuracy better than $3''$.  Uranus and Neptune were used for flux density calibration, yielding an absolute accuracy better than 10\%. 
The resulting root-mean-square (r.m.s.) noise over the mapped 4-deg$^2$ region  of the Lupus-I cloud is 5 mJy beam$^{-1}$. 

We retrieve Level-2 images of \emph{Herschel}/SPIRE from the \emph{Herschel} Science Archive (HSA).  The SPIRE data were obtained in 2011 January in parallel mode (Observation ID: 1342213182) and are processed through a pipeline software with a standard processing generation (SPG) version of v8.2.1.  The resulting 250--500 $\micron$ images cover a $2\arcdeg \times 2.3\arcdeg$ region of Lupus-I with typical $1\sigma$ noise levels of 5--15~mJy~beam$^{-1}$.  The details of the data are given by \citet{Rygl13}.

For near to mid-IR photometry, we refer to the public source catalogs obtained with the Infrared Array Camera \citep[IRAC,][]{Fazio04} and Multiband Imaging Photometer for \textit{Spitzer}  \citep[MIPS,][]{Rieke04} onboard \emph{Spitzer}, WISE \citep{Wright10} and the Two Micron All Sky Survey \citep[2MASS,][]{Skrutskie06}, which are available at the NASA Infrared Science Archive (IRSA).
The \emph{Spitzer} photometric data are provided by the \emph{Spitzer} Space Telescope ``From Molecular Cores to Planet-forming Disks'' (C2D) Legacy Program \citep{Evans03, Evans09}.  In addition, we also retrieve the basic calibrated data (BCD) from the \emph{Spitzer} Heritage Archive (SHA) in order to estimate the noise levels of the IRAC 3.6 and 5.8 $\micron$ and MIPS 24 $\micron$ photometry.  The BCD of IRAC and MIPS are processed through masking, flat fielding, background matching, and mosaicing using a standard single frame pipeline on the \textsc{Mopex} software.  The resulting r.m.s.\ noise levels at 3.6, 5.8 and 24 $\micron$ are 0.2, 1.5 $\mu$Jy~beam$^{-1}$ and 0.1 mJy~beam$^{-1}$, respectively.  The details on the \emph{Spitzer} observations and catalogs are reported in \citet{Evans09}.

The NRAO Very Large Array Sky Survey \citep[NVSS,][]{Condon98} and the Digitized Sky Survey (DSS) data are also available at the positions of MM-J1545 and MM-J1541.


\subsection{Ancillary Data for MM-J1545}

For MM-J1545, we used the Submillimeter Array \citep[SMA,][]{Ho04} to measure its precise position and spatial extent, and then subsequently obtained multi-wavelength ancillary data that we describe below.


\subsubsection{The Submillimeter Array Observations}

MM-J1545 was observed with the SMA at 890~$\micron$ in 2010 January and at 1.3~mm in 2011 April.  The 890-$\micron$ observations were performed in the extended configuration with eight antennas, which provided projected baselines ranging from 15 to 171 meters. The observing conditions were good (225-GHz zenith opacity of 0.05).  The double side-band (DSB) receivers were tuned to a local oscillator (LO) frequency of 335.15~GHz, and the correlator provided 4-GHz band width in each sideband.  The 1.3~mm observations were carried out in the compact configuration. The receivers were tuned so that $^{12}$CO, $^{13}$CO and C$^{18}$O $J$ = 2--1 emission lines from the Lupus-I cloud (i.e., $z=0$), as well as 1.3 mm continuum emission of MM-J1545, can be imaged.  The LO frequency was set to 224.86~GHz.  The atmospheric transparency was again good.  In both observing runs, two quasars J1626$-$298 ($S_{\rm 890\mu m} = 1.6$ Jy) and J1454$-$377 (0.33 Jy) were used for complex gain calibration, while the passband response was calibrated using 3C273 and 3C279.  The absolute flux density was scaled from the primary calibrator Titan.  The accuracy of the flux calibration is estimated to be 15\%.  

All of the data editing and calibration were performed using \textsc{idl}-based standard routines in the \textsc{mir} software package.  The calibrated visibility data were imaged (Fourier-transformed) and deconvolved using \textsc{miriad} \citep{Sault95} tasks, \texttt{invert} and \texttt{clean}, respectively.  In continuum imaging, the upper (USB) and lower sideband (LSB) data, eliminating channels where the local molecular lines are expected, were combined before imaging.  The natural-weighted synthesized beam sizes at 335~GHz and 225~GHz are $1\farcs 49 \times 1\farcs 17$ (the position angle, PA = $-53 \fdg 8$) and $4\farcs 94 \times 3\farcs 21$ (PA = $-5 \fdg7 $), respectively.  The resulting r.m.s.\ noise levels at 335 and 225 GHz are 3.0 and 0.76 mJy beam$^{-1}$, respectively.


\subsubsection{Karl G.\ Jansky Very Large Array Observations}

The Karl G. Jansky Very Large Array (VLA) 6~cm data were taken in C-configuration in 2010 November (project ID: 10C-226).  The correlator was configured to provide $16 \times 128$~MHz subbands covering from 4.2 to 6.1~GHz.  The data were calibrated using the VLA Calibration Pipeline,\footnote{See \url{https://science.nrao.edu/facilities/vla/data-processing/pipeline}} which is based on the \textsc{casa} data reduction package \citep{McMullin07}.  Imaging was also carried out using \textsc{casa} employing the multi-frequency synthesis algorithm (spectral Taylor expansion) with \texttt{nterms} = 2 during the deconvolution, to take into account the spectral index of the sources within the field \citep{Rau11}, along with Briggs weighting (\texttt{robust} = 0.5).  The resulting synthesized beam is $10\farcs 36 \times 3\farcs 22$ at position angle $3\fdg 7$, and the r.m.s.\ noise is 5.8~$\mu$Jy~beam$^{-1}$.


\subsubsection{The Australian Telescope Compact Array Observations}

The Australian Telescope Compact Array (ATCA) was used to take 7~mm continuum data in the H214 array configuration in 2013 October and in the H168 configuration in 2014 April (project ID: C2910).  We used five tunings to fully cover the 33.4--50.4 GHz band to search for a redshifted $^{12}$CO~(2--1) line (Taniguchi et al., in preparation).  The center frequencies were set to 35.25, 38.75, 42.20, 45.40, 48.60 GHz, and two 2 GHz spectral windows of the CABB correlator were configured adjacently, resulting in an instantaneous frequency coverage of 3.9~GHz for each tuning.  The complex gain was monitored using a radio source B1541$-$375 ($S_\mathrm{7mm} = 1.1$~Jy, $10\fdg 7$ away from MM-J1545).  Bandpass and delay calibrations were performed using 3C~279 once per night before the observations started.  Mars was used for absolute flux density calibration.  The absolute flux density uncertainty is estimated to be $<15\%$. 
The data were calibrated using \textsc{miriad} and imaged using \textsc{casa} with Briggs weighting (robust parameter of 0.5).  We did not use the five longest baselines including the antenna at the W392 station because of poor phase stability.  The resulting synthesized beam size and r.m.s.\ noise level were $6\farcs 02 \times 4\farcs 43$ (PA = $82\fdg 4$) and 35~$\mu$Jy beam$^{-1}$, respectively.


\subsubsection{Nobeyama Observations}

We also use the Nobeyama Millimeter Array (NMA) at Nobeyama Radio Observatory (NRO) to constrain the 3 mm photometry.  The observations were done during 2010 April and May in the D configuration, where only five antennas were operational.  The DSB receivers were tuned at 98.20 GHz (LSB, $\lambda = 3.05$~mm) and 110.20 GHz (USB), and the UWBC correlator \citep{Okumura00} with a 1-GHz bandwidth was used.  We did not use the USB data because of poor quality.  The visibility data were calibrated with \textsc{uvproc-ii} \citep{Tsutsumi97}, and then imaged using the \textsc{aips} \citep{Greisen03} task, \texttt{imagr}. The resulting beam size and r.m.s.\ noise level are $16\farcs 2 \times 9\farcs 4$ (PA = $-16 \fdg 3$) and 2.0 mJy beam$^{-1}$, respectively.

Since MM-J1545 is toward the edge region of a Galactic molecular cloud, it is necessary to carefully investigate whether the source is indeed extragalactic.  We used the NRO 45-m telescope to observe Galactic $^{12}$CO~(1--0), $^{13}$CO~(1--0) and C$^{18}$O~(1--0) emission lines.  The 45~m observations were performed during 2010 January to April.  The $^{12}$CO and $^{13}$CO observations were carried out with the on-the-fly (OTF) mode of the multi-beam BEARS receiver \citep{Sunada00} and with the position switching mode of the T100 single-beam receiver \citep{Nakajima08}, respectively.  The $^{12}$CO map covered a $17' \times 17'$ region including MM-J1545 and a part of the Lupus-I cloud located north-east of MM-J1545.  The $^{13}$CO OTF observations covered a $4' \times 4'$ region centered on MM-J1545.  In both of the observations, the AC45 digital spectrometer was used.  For the C$^{18}$O observations, we used the T100 single-beam receiver and the acousto-optical spectrometers with a high-dispersion mode (AOS-H) in the standard position-switching mode, providing a spectral resolution of 0.10 km s$^{-1}$ at 110 GHz.  Intensity calibration was done using the single-temperature chopper-wheel method, and the accuracy of intensity calibration is estimated to be 20\%.


\subsubsection{Near-infrared Observations}

We carried out near-infrared (NIR) imaging observations of MM-J1545 with the Subaru telescope equipped with the MOIRCS instrument \citep{Ichikawa06, Suzuki08} on 2010 April 23.  The observations were made with the $K_\mathrm{s}$-band filter at $\lambda$ = 2.15 $\micron$, with a pixel scale of $0.12''$ pix$^{-1}$.  The total integration time was 2.9~ksec. The position and magnitude were calibrated with several 2MASS point sources with $K_\mathrm{s} \sim 15$ within $\sim 1'$.  The systematic uncertainties of the astrometry and magnitude are estimated to be $0\farcs 1$ and 0.07 mag, respectively. 

\textit{JH}-band imaging observations were performed with the Wide Field Camera \citep[WFCAM,][]{Casali07} attached to the United Kingdom Infrared Telescope (UKIRT) on Mauna Kea, during 2010 March 15--19.  These observations were complemented with the MOIRCS $K_\mathrm{s}$ and \textit{Spitzer} data.  The sky was photometric throughout these nights and seeing sizes were mostly $0\farcs 7$--$1\farcs 1$.  A five-point dithering and a four-point micro-stepping were used in all observations to compensate for bad pixels and to recover full point spread function sampling.  This results in proper sampling of the seeing size with the $0\farcs 4$ WFCAM pixels.  Each exposure time was 10~sec, yielding total integration times of 2.8~ksec and 2.4~ksec at $J$ and $H$, respectively.  All of the data were reduced in a standard manner.  The astrometric uncertainties are less than $0\farcs 1$, and the photometric errors at $J$ and $H$ are 0.18 and 0.32 mag, respectively.


\begin{figure}[tbp]
	\begin{center}
		\includegraphics[scale=0.44,angle=-90]{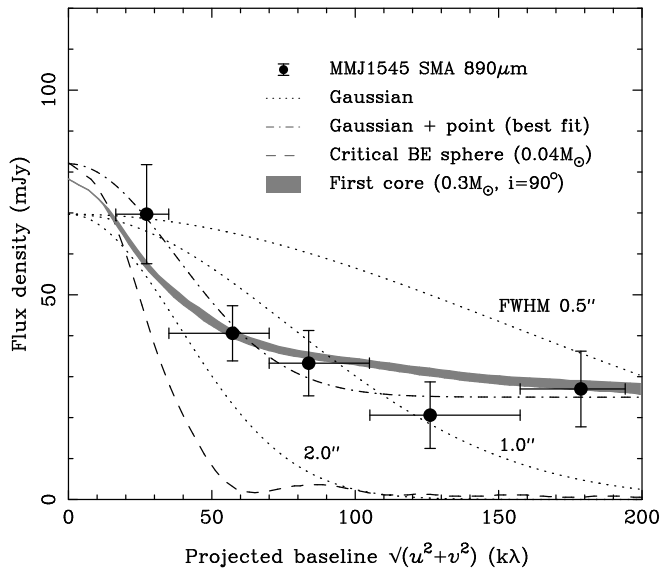}
		\caption{
			The flux densities versus projected baseline length normalized by an observing wavelength 
			(i.e., spatial frequency, $\sqrt{u^2+v^2}$) for the 890-$\micron$ SMA image of MM-J1545.
			The filled circles show the visibility amplitudes of the SMA counterpart.  
			The amplitude is well fitted by a Gaussian with a point source (dash-dotted curve), 
			suggesting an unresolved point- or cusp-like structure embedded 
			in a $\approx 2''$ extended component.  Just for reference, 
			we plot circular-symmetric Gaussian profiles with a total 890-$\micron$ flux of 
			69.7~mJy and FWHM of $0\farcs 5$, $1\farcs 0$, and $2\farcs 0$ (dotted curves).
			We show the Fourier components of a $0.04\,M_{\sun}$ Bonnor--Ebert (BE) sphere 
			\citep{Ebert55, Bonnor56} as a realization of a prestellar core.  
			The gas temperature and central H$_2$ density is assumed to be 7.1~K 
			and $5 \times 10^{7}$~cm$^{-3}$, respectively.
			We also show the predicted 890-$\micron$ Fourier distribution of 
			a first hydrostatic core with inclination angle of $i = 90\arcdeg$ collapsed 
			from a $0.3\,M_{\sun}$ BE sphere, which is computed 
			by radiation hydrodynamic simulations \citep{Tomida10,Saigo11}, 
			while we conclude that MM-J1545 is not likely a first hydrostatic core, 
			but a $z \simeq 4$--5 starburst galaxy, from multi-wavelength analysis.
			See \S~\ref{sect:fc} for details.
		}\label{fig:uvplot}
	\end{center}
\end{figure}


\section{Results}\label{sect:results}

\subsection{Individual Sources}\label{sect:individual}

MM-J1545 and MM-J1541 are detected at 1.1~mm at signal-to-noise ratios (SNR) of 7.8 and 5.4, respectively.   The flux densities of MM-J1545 and MM-J1541, which are measured for the PCA-cleaned map, are $43.9 \pm 5.6$~mJy and $27.1 \pm 5.0$~mJy.  
We do not correct for possible flux boosting due to underlying fainter sources since it is  difficult to estimate the Bayesian prior from uncertain number counts at $S_{\rm 1.1mm} > 10$~mJy.  They are, however, unlikely flux-boosted because such bright sources are extremely rare and, as we shall discuss in \S~\ref{sect:counts}, the shape of the number counts are appear to be much flatter than those at $S_{\rm 1.1mm} < 10$~mJy, where flux boosting is significant.  These two are amongst the four brightest ever discovered in the AzTEC/ASTE campaign, one of which is a $S_{\rm 1.1mm} = 37$~mJy SMG at $z = 3.39$ reported by \citet{Ikarashi11} \citep[known as \textit{Orochi} or HXMM02,][]{Wardlow13} and the other is a 43-mJy source toward the peripheral field of the Small Magellanic Cloud reported elsewhere \citep{Takekoshi13}.  Both of the sources toward the Lupus-I cloud are also detected in the SPIRE 250, 350, and 500-$\micron$ bands, most of which have flux densities of $\gtrsim$100-mJy.  In this respect, both are very similar in FIR-to-mm brightness
to the extremely-luminous SMGs detected in the \emph{Herschel}-ATLAS \citep{Gonzalez-Nuevo12} and HerMES surveys \citep{Wardlow13}.  The coordinates and results of photometry are listed in Table~\ref{tab1_Multiwavelength}.  The postage stamp images are given in Figure~\ref{fig:stamp}.

An important point that we must note is that the sources are found toward a Galactic ($z = 0$) molecular cloud.  Multi-wavelength data require an extremely cold ($< 10$~K) and compact ($\ll 10''$) nature of the sources, 
which is too rare among Galactic star-forming objects and thus unlikely associated with the molecular clouds, but are located at cosmological distances.   The spatial profiles are all consistent with a point-like source with the AzTEC ($\approx 37''$) and SPIRE (20--30$''$) beams, unlike starless cores found in Galactic molecular clouds, which are typically found to be $\sim 0.1$~pc corresponding to $\sim 100''$ at the distance to the Lupus-I cloud \citep[e.g.,][]{Onishi02}. 
The SPIRE photometry places a constraint on the peak positions of dust spectral energy distributions (SEDs) at $\lambda _{\rm obs} \gtrsim 300$--400 $\micron$, which indicates the effective dust temperatures of $T_{\rm dust} / (1+z) \lesssim 10$~K.   Given the dust temperatures found in starburst galaxies ($\gtrsim 40$~K), this suggests that MM-J1545 and MM-J1541 are situated at $z \gtrsim 3$.  We will perform photometric redshift estimates in \S~\ref{sect:photoz}.  Below is a summary of the individual sources. 


\begin{figure*}[htbp]
\begin{center}
	\includegraphics[scale=0.37,angle=-90]{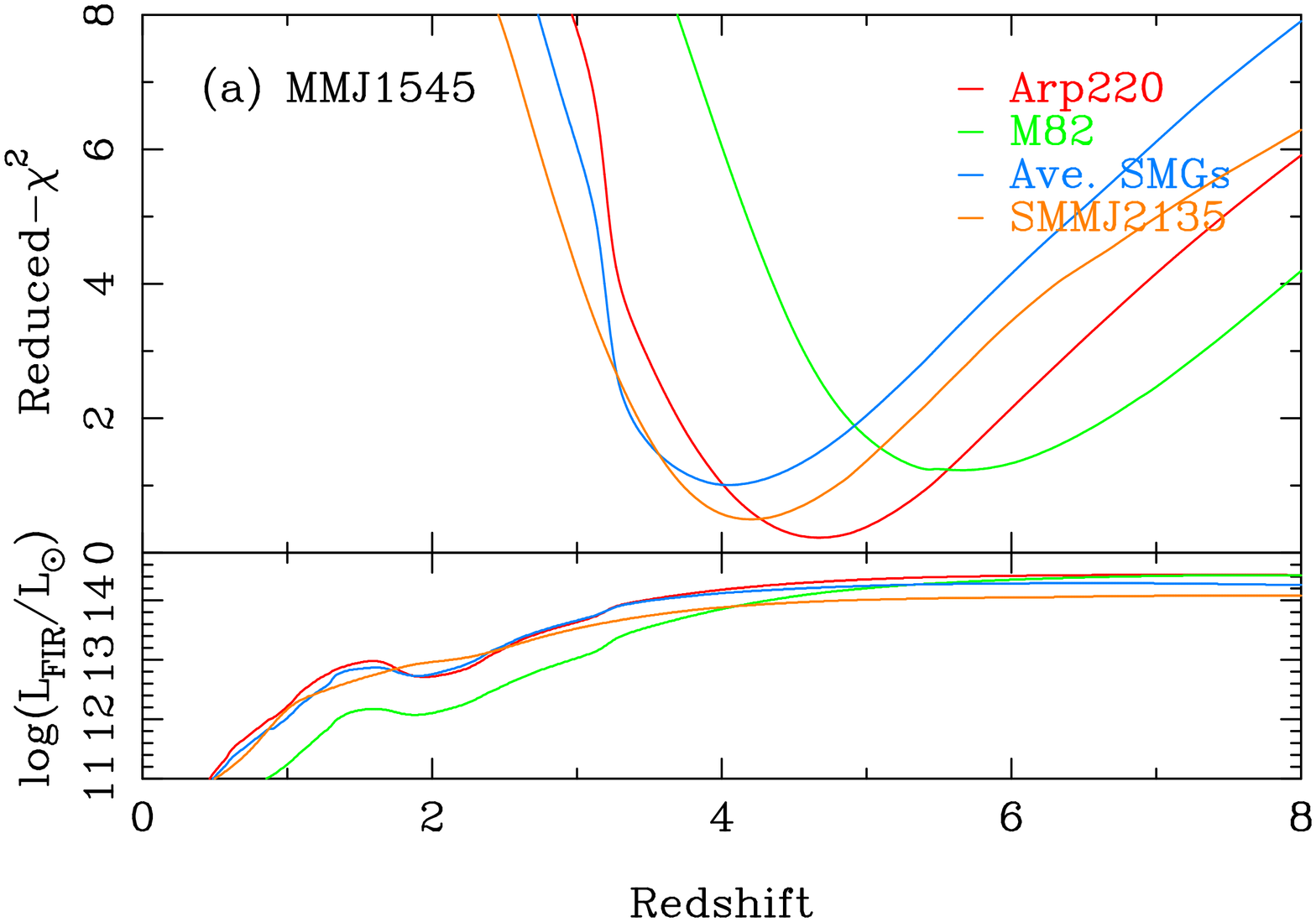}
	\includegraphics[scale=0.37,angle=-90]{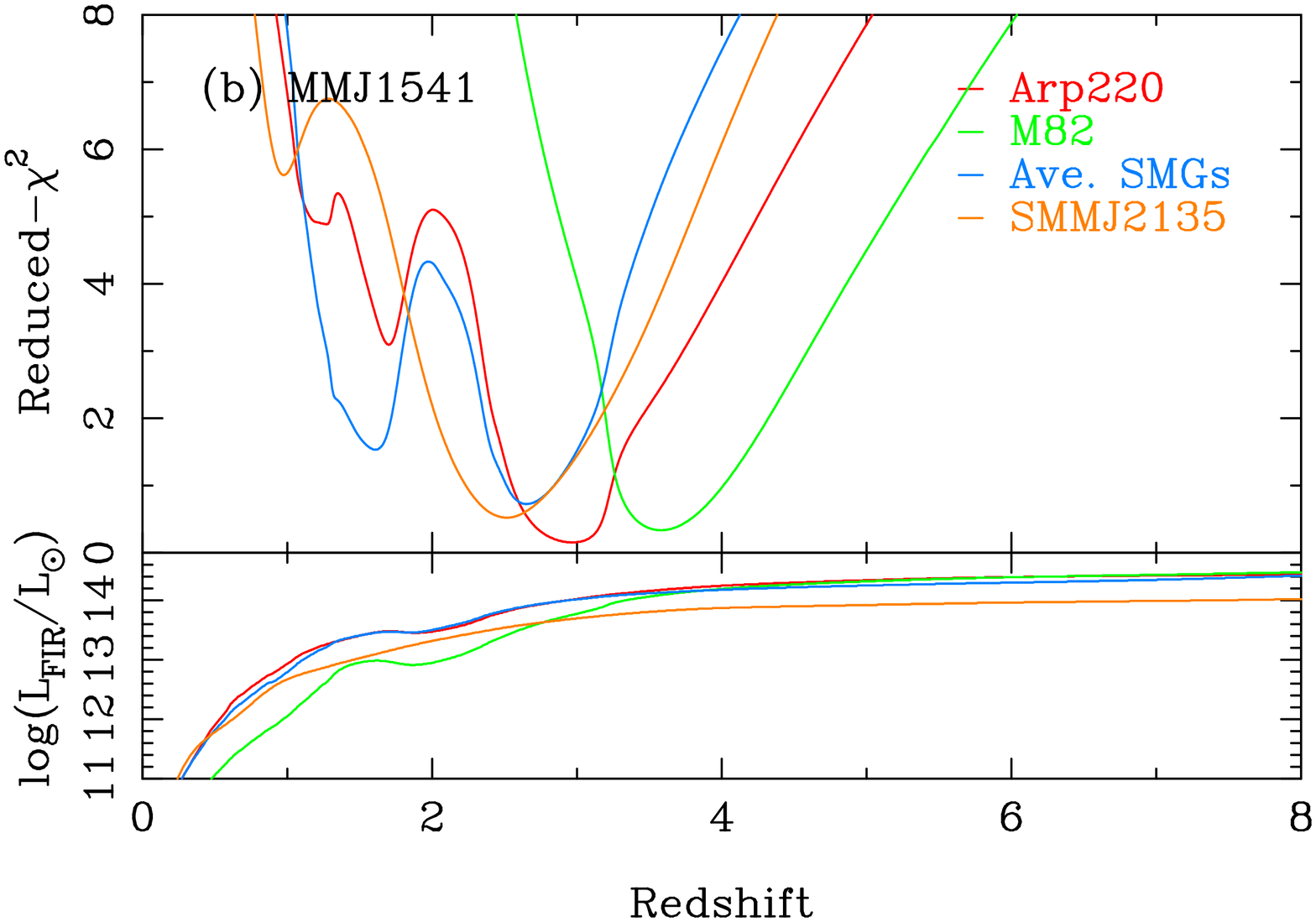}\\
	\includegraphics[scale=0.37,angle=-90]{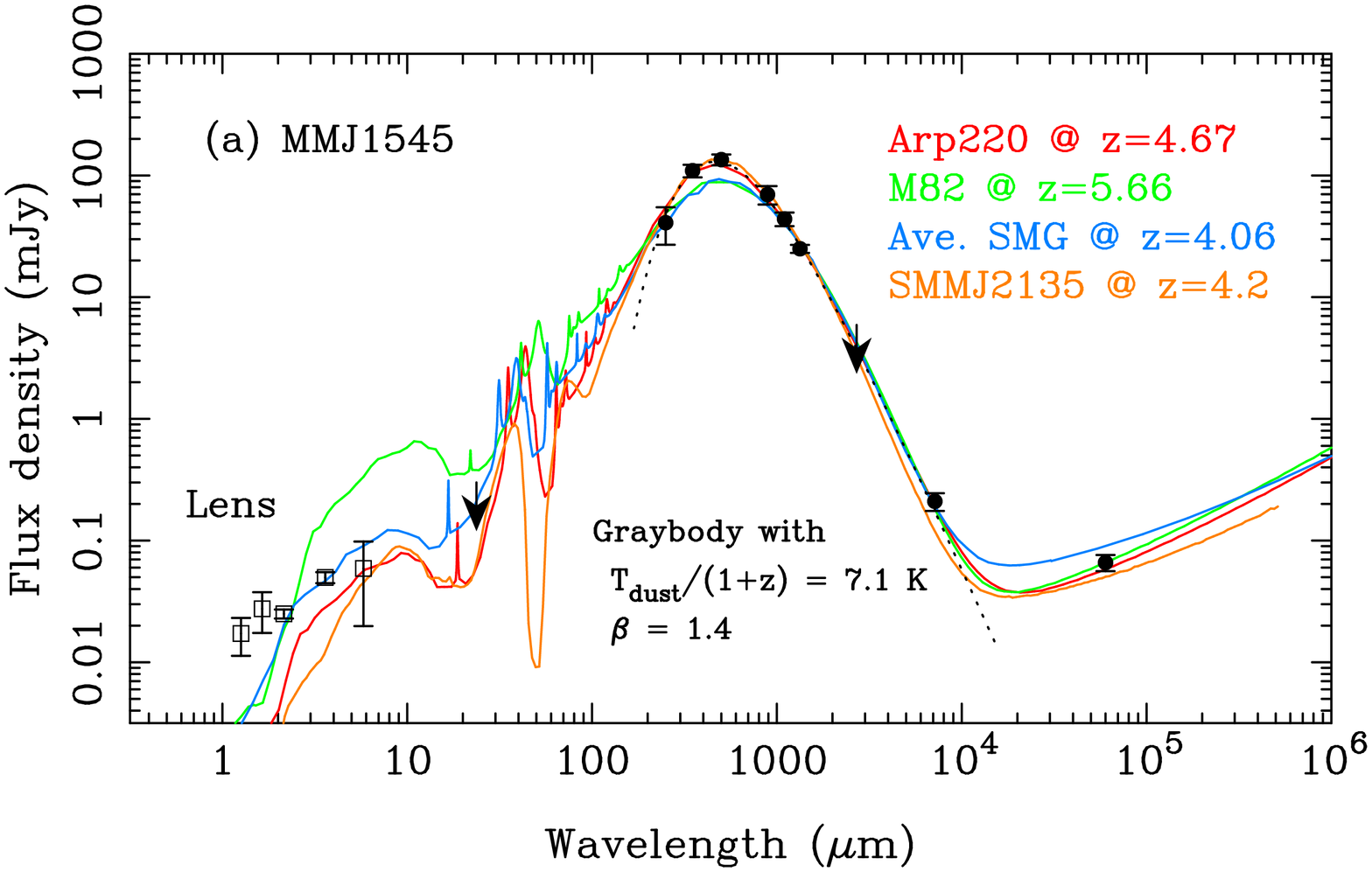}
	\includegraphics[scale=0.37,angle=-90]{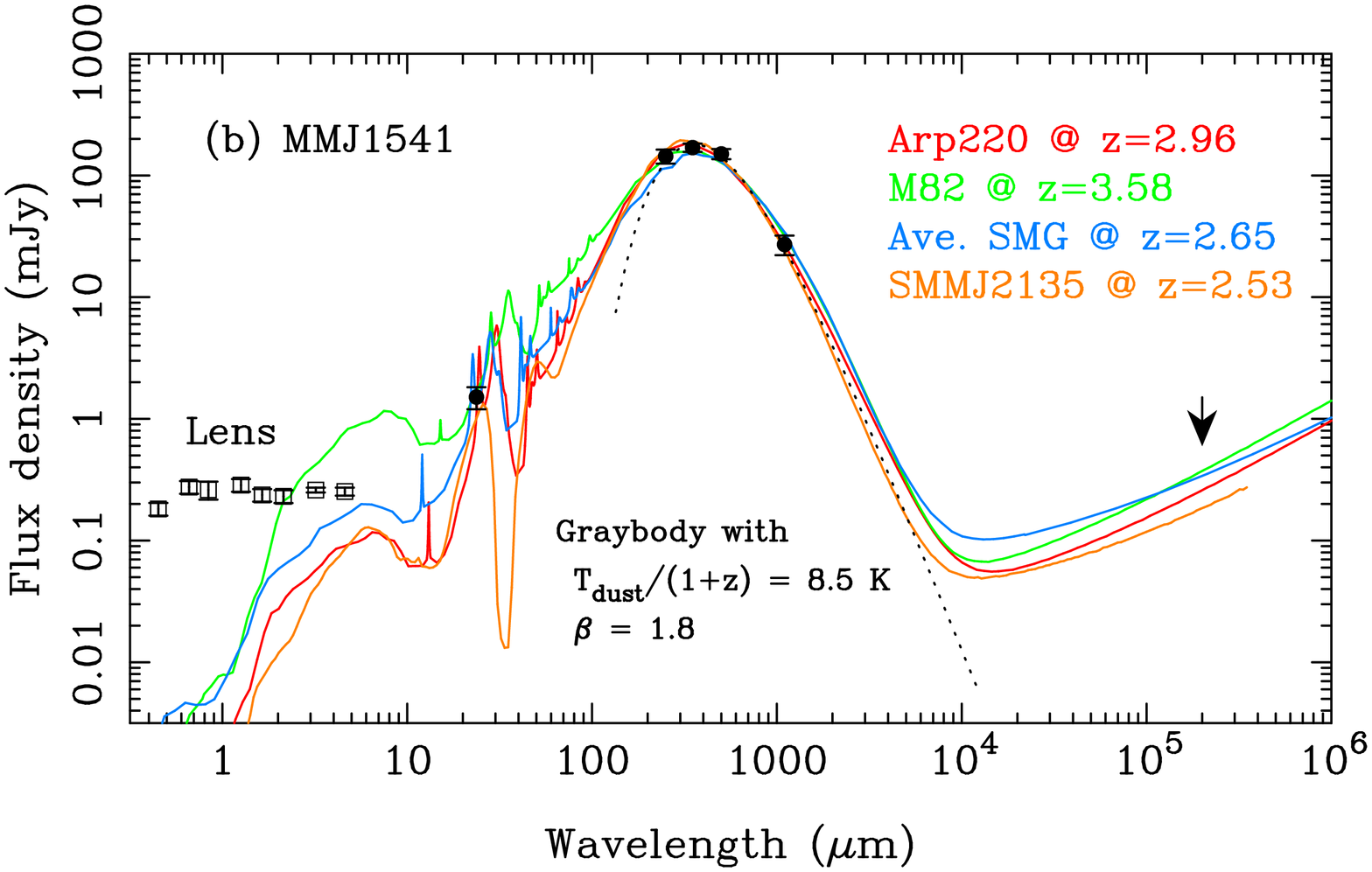}
	\caption{
		Radio-to-FIR photometric redshifts of (a) MM-J1545 and (b) MM-J1541.
		({\it top}) The upper panels show the reduced-$\chi^2$ as a function of redshift. 
		The lower panels show the far-infrared (FIR) luminosity that minimize $\chi^2$ at given redshift. 
		To avoid contamination from a (possible) foreground lensing object, 
		we only use photometric data at $\lambda _{\rm obs} \ge 24~\micron$.
		({\it bottom}) The best-fit SEDs of the targets.  
		The black circles show flux densities of the targets which used in the photo-$z$ estimates, 
		while the open squares indicate those of nearby lens candidates, J1545B and J1541B, 
		which are not used in photo-$z$ estimates.
		The dotted curves are the best-fit function of a modified black-body (gray-body).
		The measured flux densities of MM-J1545 are well fitted by a gray-body 
		with $T_{\rm dust}/(1+z) = 7.1$~K and $\beta = 1.4$, 
		while the SED of MM-J1541 is described by a $T_{\rm dust}/(1+z) = 8.5$~K 
		gray-body ($\beta$ is fixed to 1.8 
		because of the small number of photometric data points).
	}\label{fig:photoz}
\end{center}
\end{figure*}

\subsubsection{MM-J1545}\label{sect:mmj1545}

The SMA interferometric observations at 890 $\micron$ and 1.3~mm confirms the exact position at the J2000 equatorial coordinate of ($\alpha,\,\delta$) = (${\rm 15^h 45^m 6 \fs 347,\, -34\arcdeg 43' 18\farcs 18}$) with the uncertainties of $\simeq 0\farcs 09$ and $\simeq 0\farcs 10$ at 890~$\micron$ and 1.3~mm, respectively (see below for estimation of positional uncertainties).  
 The flux densities at 890~$\micron$ and 1.3~mm are $69.7 \pm 12.1$~mJy and $20.8 \pm 1.9$~mJy, respectively. The 890~$\micron$ image is resolved with its $1''$ beam, and the beam-deconvolved source size fitted with a single 2-dimensional Gaussian using a \textsc{miriad} task \texttt{uvfit} is $1\farcs 2 \times 0\farcs 64$ (PA = $37\arcdeg$).  Figure~\ref{fig:uvplot} shows the visibility amplitudes versus projected baseline length (i.e., the Fourier transform of a radial profile as a function of spatial frequency) for MM-J1545.  The Fourier components are well expressed by a single Gaussian  ($2\farcs 1 \pm 0\farcs 6$ in FWHM) with a constant offset ($25\pm 6$~mJy), implying a cusp-like compact structure embedded in a extended ($\approx 2''$) component.

At the SMA position, VLA 6~cm and ATCA 7~mm emission is also detected at $66 \pm 5~\mu$Jy and $210 \pm 35~\mu$Jy, respectively, but it is not resolved with the VLA and ATCA beams.   We do not detect 2.7 mm emission in the NMA image, which provides a $3\sigma$ upper limit of 5.9 mJy.  The counterpart was not detected in the MIPS 24~$\micron$ down to the $3\sigma$ limiting flux density of 0.3~mJy.

In Figure~\ref{fig:photoz}, we fit the SED from the infrared to the radio with a single-component modified blackbody (or graybody), $\kappa_{\rm d} B_{\nu}(\nu, T)$, where $B_{\nu}$ is the Planck function and $\kappa_{\rm d} = \kappa_{0} \nu^{\beta}$ is the dust absorption coefficient which follows a power-law function of frequency $\nu$.  The FIR-to-mm part of the SED is well described with a single modified blackbody with an effective temperature of $T_{\rm dust}/(1+z) = 7.1 \pm 0.3$~K and the emissivity index of $\beta = 1.4 \pm 0.1$.  However, the 6-cm flux clearly exceeds the gray-body function. The spectral index over the 6~cm band is $\alpha = 0 \pm 1$ (the error is the 1$\sigma$ confidence interval), where $S_{\nu} \propto \nu^{\alpha}$. This is rather flat compared with the Rayleigh-Jeans slope of $\alpha = 2 + \beta = 3.4 \pm 0.1$, suggesting that the 6-cm flux arises from synchrotron and/or free-free emission.

\begin{figure}[tbp]
	\begin{center}
		\includegraphics[scale=0.50,angle=-90]{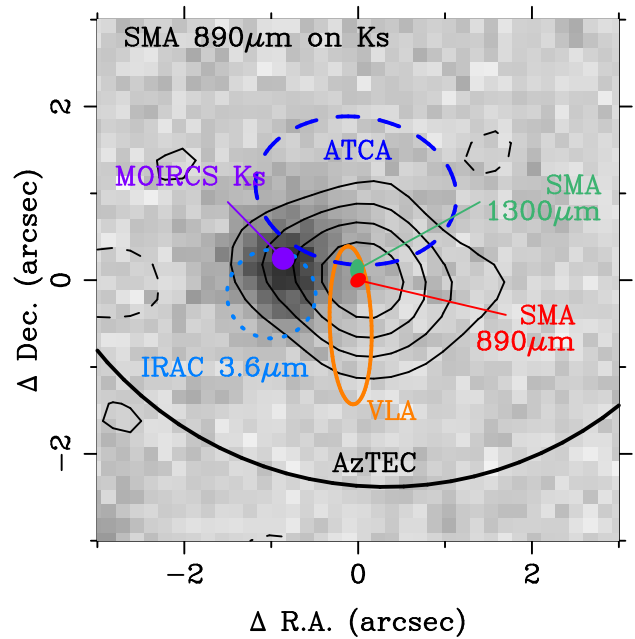}
		\caption{
			The 1$\sigma$ positional uncertainties of MM-J1545 measured at near-IR 
			(MOIRCS $K_\mathrm{s}$, purple filled circle; IRAC 3.6\,$\micron$, light blue dashed circle) 
	and submm to radio wavelengths (SMA, AzTEC, ATCA and VLA, elipses in red, black green, blue and orange).  
			The uncertainties are estimated by adding statistical and systematic positional errors in quadrature.
	The background image and contours show the MOIRCS $K_\mathrm{s}$ and SMA 890~$\micron$, respectively.
			The contours start from $2\sigma$ with an interval of $2\sigma$ and 
			the negative flux densities are represented as dashed contours.
			The separation between the NIR and the SMA/JVLA sources is significant 
			at $\gtrsim 2\sigma$ confidence level.
		}\label{fig:astrometry}
	\end{center}
\end{figure}

In the MOIRCS, WFCAM and IRAC images, a faint ($K_\mathrm{s} = 18.55$ in the Vega magnitude system, 
$S_{\rm 3.6\mu m} = 49.8 \pm 5.2~\mu$Jy) source is detected at $0\farcs 9$ east of the SMA 890-$\micron$ peak.  Hereafter we refer to this NIR object as J1545B.  Table~\ref{tab2_Lens} lists the optical to NIR photometry of J1545B.  Figure \ref{fig:astrometry} shows the centroids and uncertainties of multi-wavelength counterparts detected in the NIR (2.15~$\micron$, 3.6~$\micron$) and the submm to centimeter (890~$\micron$, 1.3~mm, 7~mm, 6~cm).  The positional uncertainty of each counterpart is estimated by adding statistical and systematic errors in quadrature.  The statistical error is obtained from $\Delta \theta _{\rm stat} \approx 0.6\,\theta_{\rm beam} / {\rm SNR}$, where $\theta_{\rm beam}$ and SNR are the beam FWHM and signal-to-noise ratio.  Given the small positional uncertainty of the SMA and MOIRCS ($< 0.1''$), this offset is significant and it is unlikely that J1545B is a counterpart to MM-J1545.   The FWHM of the MOIRCS image of J1545B is the $0\farcs 97$, whereas that of a nearby star (point spread function, PSF) is $0\farcs 69$, suggesting that J1545B is intrinsically extended (a PSF-deconvolved size of $\approx 0\farcs 68$) and likely a low-$z$ galaxy along the line of sight toward MM-J1545.  This close association of a foreground galaxy may gravitationally magnify the background object, which can naturally explain the extreme flux density of MM-J1545.  The possible gravitational lensing will be discussed in \S~\ref{sect:gravlens}
 
Furthermore, we do not detect the $J$ = 2--1 and 1--0 transitions of C$^{18}$O toward MM-J1545, as shown in Figure~\ref{fig:45obs}. From this we put a meaningful constraint on a molecular mass of a possible Galactic dense gas core, suggesting that MM-J1545 is not of Galactic origin.   The $J$ = 2--1 and 1--0 transitions of C$^{18}$O trace molecular gas with $n{\rm (H_2)} \gtrsim 10^{4.3}$ and $\gtrsim 10^{3.3}$ cm$^{-3}$, respectively, and universally seen associated with Galactic starless cores.  The $3\sigma$ upper limits on the main-beam temperature with a velocity resolution of 0.5 km~s$^{-1}$ are $T_{\rm mb} < 0.3$~K (C$^{18}$O $J=2$--1) and $T_{\rm mb} < 0.07$~K (C$^{18}$O $J = 1$--0), yielding an upper limit to the dense gas mass under the local thermodynamical equilibrium (LTE) of $< 0.005 M_{\sun}$ from the C$^{18}$O (1--0) constraint.  This LTE mass is much smaller than those found in Galactic starless cores.  No compact $^{12}$CO nor $^{13}$CO emission is significantly detected at the SMA continuum position although ambient molecular gases are contaminated across the field of view of the 45-m and SMA maps. The absence of high-velocity components in $^{12}$CO spectra, which trace molecular outflows from an accreting protostellar system, rules out any protostellar phases.

The 6~cm emission is also critical to judge if the object is extragalactic; starless cores have neither synchrotron nor free-free emission unlike galaxies.  The clear excess to the gray-body defined at 250~$\micron$ to 7~mm and the rather flat spectral index in the 6~cm band excludes the possibility that the 6~cm signal is dominated by dust emission from a Galactic starless core.  Thus, it is natural to suppose the object to be extragalactic.  We will further discuss on the Galactic possibility in \S~\ref{sect:fc}.


\subsubsection{MM-J1541}\label{sect:mmj1541}

MM-J1541 is also unlikely a Galactic source, because it is well isolated from the main clouds ($A_V > 2$) of Lupus-I and meets the ``off-cloud'' criterion defined by \citet{Rygl13}.  It is significantly detected at 24~$\micron$ with \emph{Spitzer} (SNR $\approx$ 6) at $(\alpha,\,\delta)$ = (${\rm 15^h 41^m 32\fs 706,\, -35\arcdeg 03' 19\farcs 03}$), all consistent with a $z\approx 3$ SMG, although no IRAC data are available.  A small enhancement ($2.5\sigma$, 1.3~mJy) at 20~cm is seen in the NVSS 20-cm image.   As shown in Figure~\ref{fig:photoz}, the FIR-to-mm part of the SED is well described by a single component gray-body with an effective dust temperature of $T_\mathrm{dust} / (1 + z) = 8.5$~K, where we assume a dust emissivity index of $\beta = 1.8$ because of the limited number of photometric data points.

A bright NIR/optical source is detected in 2MASS $JHK_\mathrm{s}$ and DSS $BRI$ images at $(\alpha,\,\delta)$ = (${\rm 15^h 41^m 32\fs 56}$, $-35\arcdeg 03' 23\farcs 3$), which is $3''$ south-west of the 24 $\micron$ centroid.  We hereafter refer to this 2MASS/DSS object as J1541B.  We also find a 3.2 and 4.6 $\micron$ object in WISE data at this position.  Unfortunately, we have no higher resolution images at 24 $\micron$ or longer wavelengths, at which the emission likely comes from MM-J1541.  But given that the 24 $\micron$ source is detected at high SNR and is likely a counterpart to MM-J1541, we can place a constraint on its position;  the statistical uncertainty is estimated to be $0\farcs 6$ while the systematic error is $1\farcs 4$, yielding a total positional uncertainty of 24 $\micron$ image of approximately $1\farcs 5$.  
The positions of J1541B measured in six 2MASS/DSS bands coincide with each other, and the astrometric accuracy is estimated to be better than $0\farcs 3$ for J1541B\footnote{Explanatory Supplement to the 2MASS All Sky Data Release and Extended Mission Products, \S~II.~2, http://www.ipac.caltech.edu/2mass/releases/allsky/doc/explsup.html}.  Therefore, the offset between the 24 $\micron$ peak and the 2MASS/DSS position is significant at a $\sim 2\sigma$ level.  Furthermore the shape of the SED is consistent with that of a low-$z$ passive elliptical as discussed later in \S~\ref{sect:gravlens}, suggesting that J1541B may be a galaxy which lenses the background SMG MM-J1541 at $z \sim 3$.  We list flux densities of J1541B in Table \ref{tab2_Lens} and will further discuss the possible lensing in \S~\ref{sect:gravlens}.


\subsection{Photometric Redshift Estimates}\label{sect:photoz}

We fit the SED models to the photometric data points at $\lambda_{\rm obs} \ge 24~\micron$ to constrain their photometric redshifts and FIR luminosities.  We use SED templates of well-characterized starburst galaxies;  Arp~220, M~82, \citep[GRASIL,][]{Silva98}, a composite of 76 radio-identified SMGs with spectroscopic redshifts \citep{Michalowski10}, and SMM J2135$-$0201 \citep[the cosmic eyelash,][]{Swinbank10} to search for minimum $\chi^2$ by simply varying the redshifts and luminosities of the SED templates.  We consider 20\% 
of an absolute flux density uncertainty for all photometric points in addition to the statistical error.  We also take into account the $1\sigma$ confusion noise \citep{Nguyen10} for photometric errors of the SPIRE bands. 

Figure~\ref{fig:photoz} shows the results of photometric redshift estimates and the best-fit SEDs.  The inferred redshift of MM-J1545 is $z \simeq 4$--5, although the derived photometric redshifts depends on the templates.  Overall, the measured SED is in good agreement with the Arp~220 and SMM~J2135 templates although the fit with M~82 is poorer than the others.  The best-fit redshifts and 68\% 
confidence intervals are $z = 4.67^{+0.88}_{-0.74}$ (Arp~220), $5.66^{+1.17}_{-0.87}$ (M~82), $4.06^{+0.92}_{-0.11}$ (average-SMG) and $4.20^{+0.87}_{-0.64}$ (SMM~J2135).  The inferred FIR luminosities for the best-fit redshifts are then $\log{(L_{\rm FIR}/L_{\sun})} = 14.3^{+0.1}_{-0.2}$ (Arp~220),  $14.3^{+0.1}_{-0.2}$ (M~82), $14.1^{+0.1}_{-0.2}$ (average-SMG), and  $13.9^{+0.1}_{-0.2}$ (SMM~J2135).  The dust temperature is $T_{\rm dust} \simeq 36$--40~K if $z \simeq 4.1$--4.7, which is similar to those found in SMGs \citep[e.g.,][]{Kovacs06}.

The redshift of MM-J1541 is estimated to be $z \simeq 3$, although the available photometric data are very limited.  Among the templates, Arp~220 better reproduces the actual SED than the others, and the 24 $\micron$ detection is well accounted for by the 7.7-$\micron$ feature of polycyclic aromatic hydrocarbons (PAHs).  The photometric redshifts obtained the templates are $z = 2.96^{+0.33}_{-0.40}$ (Arp~220), $3.58^{+0.54}_{-0.34}$ (M~82), $2.65^{+0.40}_{-0.27}$ (average-SMG) and $2.53^{+0.49}_{-0.42}$ (SMM~J2135).  The inferred FIR luminosities for the best-fit redshifts are then $\log{(L_{\rm FIR}/L_{\sun})} = 14.0^{+0.1}_{-0.2}$ (Arp~220),  $14.1^{+0.1}_{-0.1}$ (M~82), $13.9^{+0.1}_{-0.2}$ (average-SMG), and  $13.6^{+0.1}_{-0.2}$ (SMM~J2135).  If we consider $z \simeq 3.0$, then the dust temperature is approximately 34~K, again consistent with those found in SMGs.


\begin{figure*}[htbp]
	\begin{center}
		\includegraphics[angle=-90,width=0.99\textwidth]{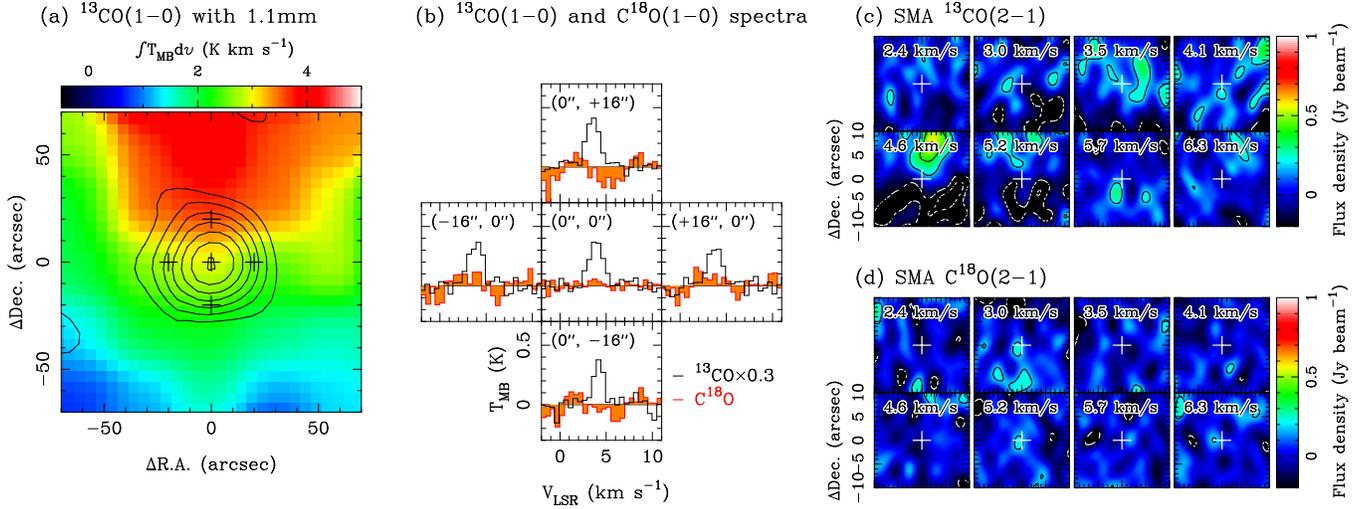}
		\caption{
			Results from the 45~m and SMA observations of Galactic molecular gas 
			along the sight line of MM~J1545.
			(a) The 45-m integrated intensity map of $^{13}$CO~(1--0) overlaid 
			with the AzTEC 1.1-mm image (contours). The contours start at 
			$1\sigma$ (5.6 mJy~beam$^{-1}$) with an interval of $1\sigma$.  
			The crosses mark the positions where C$^{18}$O~(1--0) data were obtained.
			(b) The $^{13}$CO~(1--0) and C$^{18}$O~(1--0) spectra obtained with the 45~m telescope.  
			The  $^{13}$CO spectra are scaled by 0.3$\times$ for clarity.  
			All of the spectra are shown in main-beam temperature scale $T_{\rm mb}$ and 
			have a velocity resolution of 0.5~km~s$^{-1}$.  With the 45-m beam (16$''$), 
			$^{13}$CO~(1--0) is clearly seen at the position of MM~J1545, 
			but the spectra at (0$''$, 0$''$) is consistent with the four adjacent spectra, 
			suggesting that the $^{13}$CO emission does not come from MM~J1545.  
			No C$^{18}$O emission is found.
			(c) (d) The channel maps of $^{13}$CO~(2--1) and C$^{18}$O~(2--1) 
			obtained with the SMA.  The images are not cleaned.  
			Crosses represent the position of the SMA 890 $\micron$ counterpart.  
			The velocities in terms of the local standard of rest (LSR) are indicated 
			at the top of each panel.  The contours are drawn at 
			($-4$, $-2$, 2, 4, ...)$\times \sigma$, where $\sigma \simeq 0.1$~Jy~beam$^{-1}$ 
			with a velocity resolution of 0.55~km~s$^{-1}$.  
			No compact emission is detected in $^{13}$CO~(2--1) and C$^{18}$O~(2--1).
		}\label{fig:45obs}
	\end{center}
\end{figure*}


\section{Discussions}\label{sect:discussions}

\subsection{Possibility of Galaxy--Galaxy Lensing}\label{sect:gravlens}

The extreme luminosities of MM-J1545 and MM-J1541 are likely attributed to galaxy-galaxy lensing;  In many cases, an elliptical galaxy seen in the optical to NIR is associated with brightest ($L_{\rm FIR} \ge 10^{14} L_\sun$) SMGs.  Interferometric imaging of these bright sources often reveals multiply split images or extended structures that are well explained by strong gravitational lensing models \citep[e.g.,][]{Negrello10, Ikarashi11, Wardlow13, Vieira13, Bussmann13}.  In addition, detailed modeling of the mm/submm source number counts suggests that the excess of the counts at high flux densities ($S_{\rm 1.1mm} > 10$~mJy for example) are dominated by strongly lensed SMGs as well as nearby galaxies \citep[][]{Negrello10}. 

The NIR sources, J1545B and J1541B, may be lensing objects.  While the lack of multiband optical photometry in J1545B makes it difficult to determine the lensing properties, the clear 2MASS/DSS detections of J1541B allow us to estimate the magnification factor and thus the intrinsic nature of MM-J1541.   Hence, we focus on the lensing property of the MM-J1541--J1541B system first, and then discuss the MM-J1545--J1545B system.


\subsubsection{MM-J1541}

\begin{figure}[htbp]
	\begin{center}
		\includegraphics[angle=-90,width = 0.47\textwidth,keepaspectratio,clip]{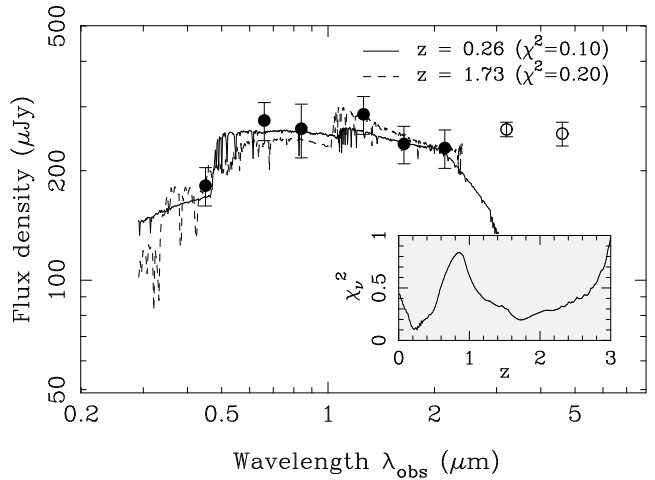}
		\caption{
			The optical to near-infrared spectral energy distribution (SED) of J1541B.  
			We fix a Galactic extinction of $E(B-V) = 0.3$, which approximately corresponds to 
		$A_V = 1$.  The photometry from WISE (3.2 and 4.6 $\micron$, open symbols) is not used   
			because the WISE bands can be affected by emission from small grains and 
			polycyclic aromatic hydrocarbons of J1541B.  The stellar emission from 
			the background MM-J1541 can also contribute to the WISE photometry.  
			The solid curve represents the best-fit SED at $z = 0.26^{+0.29}_{-0.13}$ (68\% 
			confidence interval).  The dashed curve shows the SED at the secondary 
			$\chi^2$ minimum ($z = 1.73^{+0.77}_{-0.52}$). The inset panel shows the reduced 
			$\chi^2$ as a function of redshift $z$.
		}\label{fig:sed1541b}
	\end{center}
\end{figure}

To characterize the properties of J1541B, we perform SED fits to the optical-to-NIR photometric data of J1541B using the \textsc{Hyperz} code\footnote{http://webast.ast.obs-mip.fr/hyperz/} developed by \citet{Bolzonella00}.  In SED fits, we used SED templates of \citet{Bruzual03}.  From the SPIRE $A_V$ map, the visual extinction toward MM-J1541 is approximately 1.5, but the stray light from the Moon in the SPIRE observations \citep{Rygl13} can bias against low visual extinction.  We assume a conservative value of $A_V = 1$.  Note that this assumption does not dramatically affect the result because the total extinction is considered by combining the Galactic reddening and the intrinsic extinction, which eventually compensates the uncertainty in the Galactic extinction.  Figure~\ref{fig:sed1541b} shows the result of SED fits.  The photometric redshift (photo-$z$) of J1541B is $z_{\rm L} = 0.26^{+0.29}_{-0.13}$ ($\chi^2_{\nu} = 0.10$, the error bar is from the 68\% 
confidence interval).  The optical to NIR SED is well described by a 33~Myr elliptical model with a stellar mass of $1 \times 10^{11}\,M_{\sun}$ and an intrinsic extinction of $A_V = 0.10$.  The $B$-band decrement can be accounted for by the 4000 \AA\ break at $z \sim 0.3$, which is well within the photo-$z$ range.  The best-fit stellar age is young (33 Myr) but maturer stellar SEDs can reasonably match the actual SED as well.

Then we make use of the Faber-Jackson relation \citep[FJR,][]{Faber76} to constrain the velocity dispersion of the lensing galaxy, which allows us to model a singular isothermal sphere (SIS) as a lensing dark halo.  FJR has originally been proposed as an empirical relation between $B$-band absolute magnitude and velocity dispersion of galaxies.  Similar relations are now confirmed in many filter bands out to the NIR \citep{Pahre98, LaBarbera10}.  In our case, the reddening can be significant because of the foreground Galactic molecular cloud, so we use the NIR version of FJR presented in Eq.~5 of \citet{LaBarbera10}.
The Einstein radius of a SIS mass distribution is expressed as 
\[
\theta _{\rm E} = 1\farcs 154 \times 
\left( \frac{\sigma_v}{\rm 200\,km\,s^{-1}} \right)^2
\left( \frac{D_{\rm A}^{\rm S}}{\rm 1\,Gpc} \right)
\left( \frac{D_{\rm A}^{\rm LS}}{\rm 1\,Gpc} \right)^{-1},
\]
where $\sigma_v$ is the velocity dispersion, $D_{\rm A}^{\rm S}$ and $D_{\rm A}^{\rm LS}$ are the angular diameter distances from the observer to the background source and from the foreground lens to the background source, respectively.  In the redshift range of $0.1 < z_{\rm L} < 0.7$, which covers the 68\%
redshift confidence interval of J1541B, we find the velocity dispersion ranging $100 < \sigma_v < 250$ km s$^{-1}$ from the $K_\mathrm{s}$ magnitude and FJR.  When we consider the background SMG at $z = 3.0$ (\S~\ref{sect:photoz}), then we find $\theta_{\rm E} \sim 0\farcs 3$--$2''$, which is smaller than the actual separation between J1541B and the 24 $\micron$ peak.  We estimate the magnification factor $\mu_{\rm g}$ at the 24 $\micron$ position ($3''$ apart from the lens centroid) using a gravitational lensing model \textsc{glafic} \citep{Oguri10}, in which we employ a circular-symmetric SIS with the velocity dispersion $\sigma_v$ derived above.  For $z_{\rm L} = 0.26$ and $z_{\rm S} = 3.0$, we find $\theta_{\rm E} = 0\farcs 7$ and the magnification factor at the position of MM-J1541, $\mu_{\rm g} = 1.2$.  If we go to $z_{\rm L} = 0.55$ that is the edge of the 68\% 
confidence interval, then we have $\theta_{\rm E} = 1\farcs 2$ and $\mu_{\rm g} = 1.4$.  To get  the Einstein radius closer to the 3$''$ separation angle, it would be necessary to double the angular diameter distance to the lens and/or increase the luminosity of the lens galaxy by a factor of $\approx 4$, which is very difficult to achieve within the uncertainties in the measured quantities.  We have another likelihood peak at $z_{\rm L} = 1.73^{+0.77}_{-0.52}$ ($\chi^2_{\nu} = 0.20$, see Figure~\ref{fig:sed1541b}), but this is unlikely because the inferred stellar mass is too high ($1\times 10^{13}\, M_{\sun}$).  Even if this would be the case, the magnification still remains a moderate value of $\mu_{\rm g} = 2.0$.   Consequently, the magnification factor of MM-J1541 is not likely as high as $\approx 10$ but rather moderate ($\mu_{\rm g} \approx 1.2$), suggesting that MM-J1541 might be an intrinsically hyper-luminous star-forming galaxy with a demagnified 1.1-mm flux density of $\sim$20 mJy or the intrinsic FIR luminosity of $\log{(L_\mathrm{FIR}/L_{\sun})} \sim 13.7$--13.9.


\subsubsection{MM-J1545}

On the other hand, amplification for MM-J1545 at $z \simeq 4$--5 may be larger although measurement of its magnification using existing data is difficult.  The Fourier analysis of the SMA visibility data of MM-J1545 clearly shows its extended morphology (FWHM $\sim 1''$--$2''$, see Figure~\ref{fig:uvplot}), suggesting that the source could be magnified by a foreground galaxy seen as J1545B. 
Unfortunately, we only have the limited number of photometric data points for J1545B, which likely suffer from relatively large Galactic reddening of $A_V \simeq 2.4$.  We find no apparent spectral break at $\lambda_{\rm obs} \ge 1.2~\micron$, suggesting the lens redshift of $z_{\rm L} < 2$.  So we assume several lens redshifts and a source redshift ($z_{\rm S} = 4.8$, from \S~\ref{sect:photoz}) to estimate the Einstein radius in the same manner as MM-J1541.  Note that a rest-frame $K_\mathrm{s}$-band extinction of 0.2 mag is used to correct the Galactic reddening.  We find $\theta_{\rm E} > 1\farcs 5$ at $z_{\rm L} > 0.7$, which is inconsistent with our SMA image showing the smaller separation ($0\farcs 9$) and no counter-images split by a gravitational lens.  This suggest the lens redshift of $z \lesssim 0.6$.  Actually, if the lensing galaxy is at $z_{\rm L} = 0.5$, we have the absolute magnitude in rest-frame $K_\mathrm{s}$-band of $M_{K_\mathrm{s}} \sim -25$ and $\sigma_v \sim 160$--170 km s$^{-1}$ from the FJR.  This yields the Einstein radius of $\approx 0\farcs 9$, consistent with the observed situation.  

Unfortunately, we cannot exactly predict the magnification factor of MM-J1545; the SMA 890-$\micron$ image exhibits neither multiple counter-images nor a large arc/ring, and the spatial extent of the source is unknown.  Furthermore, the accurate redshifts of the lens and source are not available.  All of the facts prevent us from precisely constructing a lens model.  However, given that no lensed source with a magnification of $\mu_{\rm g} > 10$ showing a single image with a $1''$-beam has been reported thus far \citep[e.g.,][]{Bussmann13}, it should be reasonable to assume $\mu_{\rm g} \sim 10$ as an upper limit for the magnification factor of MM-J1545.  In this case, the intrinsic FIR luminosity is still very high ($L_{\rm FIR}^{\rm int} \gtrsim 1 \times 10^{13}\,L_{\sun}$), even after correcting for magnification.  Such a starburst galaxy in the hyperluminous regime at $z \simeq 4$--5 is still rare compared to existing studies \citep[e.g.][]{Riechers10, Walter12, Combes12, Vieira13, Weiss13} and is a unique laboratory to investigate properties of star-formation in the early Universe.


\begin{figure*}[htbp]
	\begin{center}
		\includegraphics[width = 0.52\textwidth,keepaspectratio,clip,angle=-90]{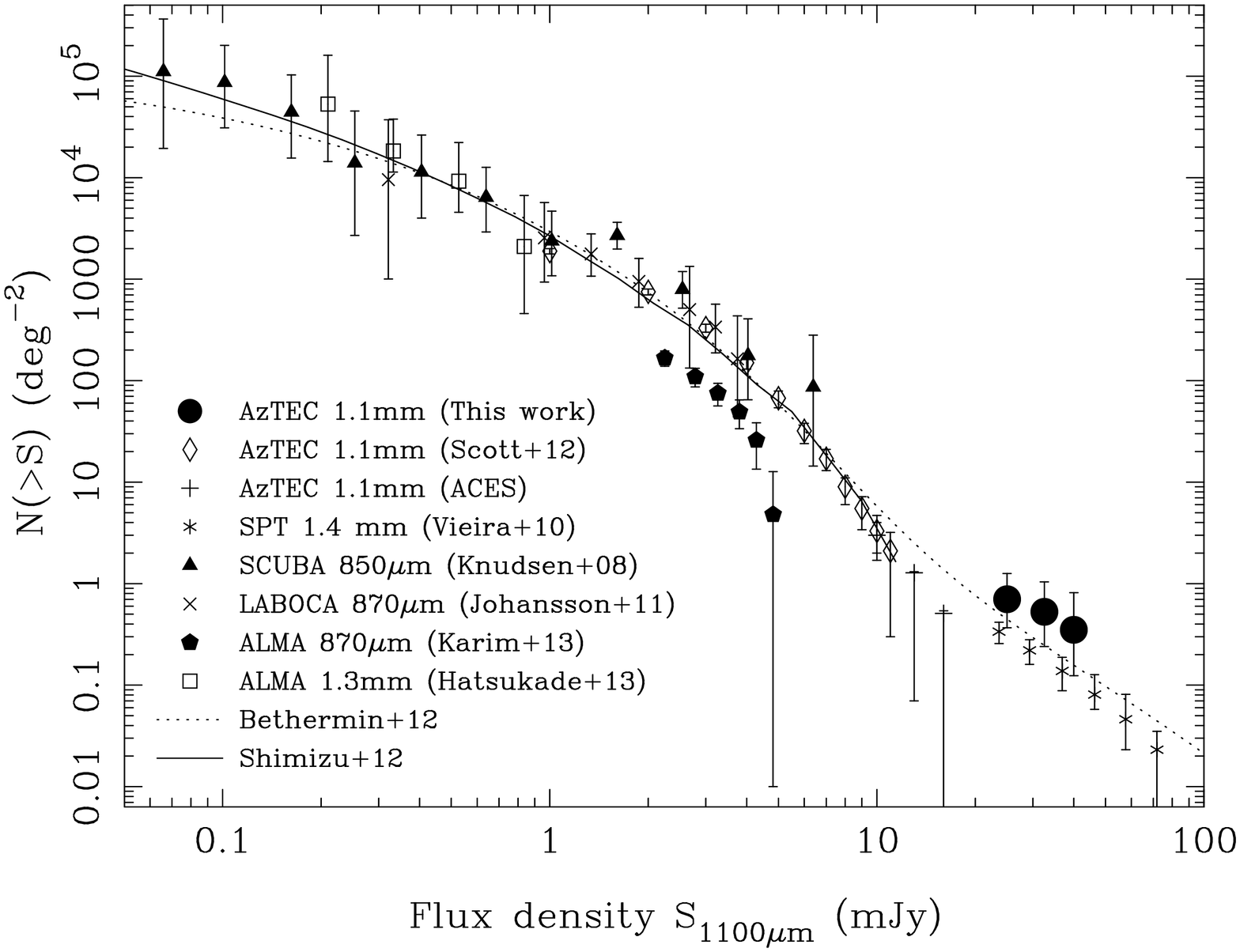}
		\caption{
			The 1.1-mm cumulative number counts obtained in the AzTEC surveys 
			(filled circles for this work, diamonds for blank field surveys from \citet{Scott12}, 
			crosses for the AzTEC Cluster Environmental Survey from \citet{Scott12}).  
			For comparison, we plot the SPT 1.4-mm counts \citep{Vieira10}, SCUBA 850-$\micron$ 
			counts \citep{Knudsen08}, LABOCA 870-$\micron$ counts \citep{Johansson11}, 
			ALMA 870-$\micron$ \citep{Karim13} and 1.3-mm counts \citep{Hatsukade13}.  
			We also show model predictions from \citet{Shimizu12} and \citet{Bethermin12}, 
			the latter of which accounts for the strong-lensing effects.  
			The 1.4-mm, 1.3-mm, 870-$\micron$, and 850-$\micron$ counts are scaled 
			to an equivalent 1.1-mm flux density using scaling factors of
			$S_{\rm 1.1mm} / S_{\rm 1.4mm} = 1.89$, 
			$S_{\rm 1.1mm} / S_{\rm 1.3mm} = 1.41$, 
			$S_{\rm 1.1mm} / S_{\rm 870\mu m} = 0.54$, and
			$S_{\rm 1.1mm} / S_{\rm 850\mu m} = 0.51$, respectively.
		}
	\label{fig:counts}
	\end{center}
\end{figure*}


\subsection{The 1.1-mm Number Counts at the Brightest End}\label{sect:counts}

The detections of ultra-bright sources allow us to constrain the brightest end of the 1.1~mm number counts, which is complementary to the deep number counts at $S_{\rm 1.1mm} = 1$--20~mJy obtained from our own 1.1-mm surveys of SMGs \citep{Hatsukade11, Scott12}.  
The area where the $1\sigma$ sensitivities of the Lupus-I AzTEC map are below 7~mJy~beam$^{-1}$ (typically 5~mJy~beam$^{-1}$) is 3.65~deg$^2$.  We eliminate the high column region where $A_V > 1$~mag on the 2MASS extinction map \citep{Dobashi11}, which leaves 2.88~deg$^2$.  We detect three $\ge 5\sigma$ point sources over the 2.88~deg$^2$ area (Tsukagoshi et al., in preparation).  
We carefully cross-identify known starless and protostellar cores \citep{Rygl13}, which leaves only two extragalactic sources, MM-J1545 and MM-J1541.  
The inferred cumulative number counts $N(S_{\rm 1.1mm} > 25$~mJy) for Lupus-I are $0.69 ^{+0.92}_{-0.45}$~deg$^{-2}$ \citep[the error is taken from the $1\sigma$ confidence interval computed by][]{Gehrels86}.   

Furthermore, two additional bright sources \citep[$S_{\rm 1.1mm} = 37.3 \pm 0.7$~mJy and $43.3 \pm 8.4$ mJy,][]{Ikarashi11, Takekoshi13} have been found to date over the AzTEC survey fields \citep[1.60~deg$^2$,][]{Scott12} and the SMC peripheral field \citep[1.21~deg$^2$,][]{Takekoshi13}, respectively.  Both of them are also indicative of strongly-lensed magnification.  Taking into account all of these four brightest extragalactic sources with $S_{\rm 1.1mm} > 25$~mJy, we estimate the cumulative number counts at $> 25$, 32.5, 40~mJy are $0.70 ^{+0.56}_{-0.34}$, $0.53 ^{+0.51}_{-0.29}$, and $0.35^{+0.46}_{-0.23}$ deg$^{-2}$, respectively, where the $1\sigma$ errors are again taken from \citet{Gehrels86}.  

Figure~\ref{fig:counts} shows the 1.1-mm cumulative number counts of the brightest AzTEC sources, as well as the deeper 1.1-mm number counts obtained from blank fields toward the Great Observatories Origins Deep Survey (GOODS) North and South fields, the Lockman Hole, the Cosmic Evolution Survey (COSMOS), the Subaru/\textit{XMM--Newton} Deep Field, the Akari Deep Field-South \citep{Scott12} and from the AzTEC Cluster Environmental Survey (ACES) \citep{Wilson08b, Scott12}.  These counts intersect at $S_{\rm 1.1mm} \sim 20$ mJy, but the slope at $S_{\rm 1.1mm} \gtrsim 20$~mJy is shallower than the deeper part of the source counts.   We also plot  the 1.4-mm counts obtained by the South Pole Telescope (SPT) survey \citep{Vieira10, Mocanu13} scaled to equivalent 1.1-mm flux densities using a scaling factor of $S_{\rm 1.1mm} / S_{\rm 1.4mm} = 1.89$ \citep{Scott12}.  The SPT counts do not include any nearby galaxies discovered by the Infrared Astronomical Satellite (IRAS) and those with synchrotron-dominated SEDs.  Alternatively, these are thought to be dominated by strongly-lensed SMGs.   The amplitude and slope at the brightest end of the 1.1-mm counts are consistent with those of the SPT counts given the uncertainty in the scaling factor.  Furthermore, a model prediction by \citet{Bethermin12}, where flux magnification due to strong gravitational lensing is accounted for, reproduces the 1.1-mm counts at $S_{\rm 1.1mm} > 25$~mJy, providing more supporting evidence that the brightest sources found at 1.1 mm are mostly attributed to the strong-lensing effect.


\begin{figure*}[htbp]
	\begin{center}
		\includegraphics[scale=0.70,angle=-90]{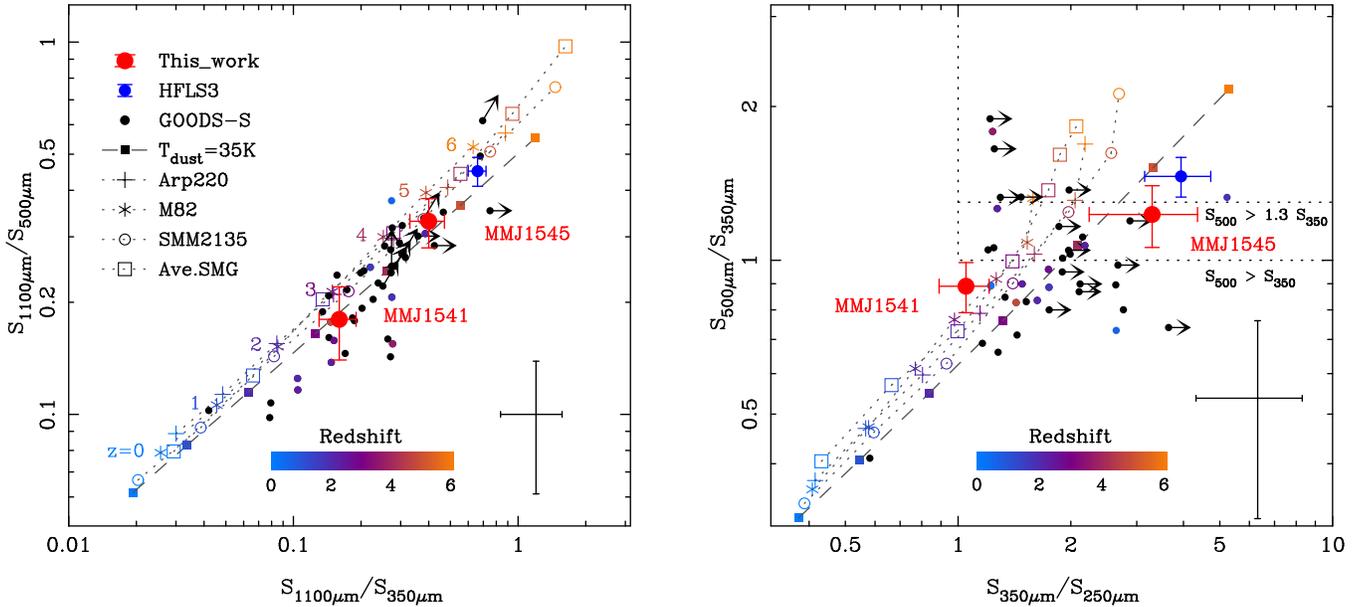}
		\caption{
			(Left) The $S_{\rm 1.1mm}/S_{\rm 350\mu m}$ versus $S_{\rm 1.1mm}/S_{\rm 500\mu m}$ 
			color-color diagram for MM-J1545 and MM-J1541 (large filled circle with error bars), 
			along with 1.1-mm selected SMGs found in the GOODS-South field (dots).  
			We also show predicted colors of template SEDs redshifted from $z=0$ to 6 
			(colored in blue to orange); a gray-body with $\beta = 1.8$ and a dust temperature 
			of $T_{\rm dust} = 35$~K (filled squares), Arp~220 (crosses), M~82 (asterisks), 
			SMM~J2135$-$0201 \citep[open circles,][]{Swinbank10}, and a mean SMG 
			\citep[open squares,][]{Michalowski10}.  The SMGs of GOODS-South with 
			spectroscopic redshifts are colored in the same way according to their redshift.  
			The typical error bar of the GOODS-South flux density ratios is shown at the bottom 
			right corner, which is mostly dominated by confusion noises in the SPIRE image.  
			For reference, we also plot the flux density ratios for the most distant SMG known, 
			HFLS3 at $z = 6.34$ \citep[blue filled circle,][]{Riechers13}.
			(Right) The \emph{Herschel}/SPIRE $S_{\rm 350\mu m}/S_{\rm 250\mu m}$ versus 
			$S_{\rm 500\mu m}/S_{\rm 350\mu m}$ color-color diagram for MM-J1545 and MM-J1541.  
			We also plot 1.1-mm selected SMGs found in the GOODS-South field with 350 and 
			500~$\micron$ detections.  The symbols and colors represent the same as the left panel.  
			The region surrounded by the straight lines that satisfies 
			$S_{\rm 350\mu m}/S_{\rm 250\mu m} > 1$ and 
			$S_{\rm 500\mu m}/S_{\rm 350\mu m} > 1$ 
			shows the criteria for the 500-$\micron$ peakers, i.e., 
			$S_{\rm 250\mu m} < S_{\rm 350\mu m} < S_{\rm 500\mu m}$.  
			We also plot the region where 
			$S_{\rm 250\mu m} < S_{\rm 350\mu m} < S_{\rm 500\mu m}/1.3$, 
			in which $z \sim 6$ SMGs may fall.
		}\label{fig:ccd}
	\end{center}
\end{figure*}


\subsection{The AzTEC--SPIRE Colors}\label{sect:color}

It is increasingly becoming clear that mm to long-submm observations are more likely to select higher redshift SMGs than FIR and short-submm as predicted earlier \citep{Blain93}, and recent studies of ultra-bright sources selected in the mm to long submm actually reveal a lot of SMGs at redshift $z > 4$ out to $z \simeq 6$ \citep{Vieira13,Weiss13,Boone13}.  It has been also suggested that selecting red objects in SPIRE bands whose SEDs are peaked at the 500-$\micron$ band (i.e., objects that follows $S_\mathrm{500\mu m} > S_\mathrm{350\mu m} > S_\mathrm{250\mu m}$; so-called 500-$\micron$ peakers) is also a useful way to pick up high-$z$ candidates even at $z \sim 6$ \citep{Riechers13,Dowell14}.  MM-J1545 is estimated to be at $z \simeq 4$--5 and is formally consistent with a 500-$\micron$ peaker, implying that 1.1-mm selection and (sub)mm color investigation are quite useful for isolating $z \gtrsim 4$ SMGs.

In the left side of Figure~\ref{fig:ccd}, we plot the $S_{\rm 1.1mm} / S_{\rm 350\mu m}$ and $S_{\rm 1.1mm} / S_{\rm 500\mu m}$ flux density ratios of MM-J1545 and MM-J1541.  For comparison, we also show the same color--color plots of 48 AzTEC-selected sources with S/N $\ge 4.0$ from the AzTEC/ASTE GOODS-South survey \citep{Scott10, Downes12}, twelve of which have robust counterparts with spectroscopic redshifts ranging from $z = 0.037$ to 4.76 \citep{Yun12}.  The \emph{Herschel} data are retrieved from the HSA and the $1\sigma$ noise levels are $\simeq 0.5$--0.6~mJy~beam$^{-1}$ in all the SPIRE bands.  The detection thresholds are set to $2\sigma_{\rm faint}$ = 7.6, 9.2, 10.4~mJy~beam$^{-1}$ at 250, 350, and 500~$\micron$, respectively, where $\sigma_{\rm faint}$ is an underlying confusion limit after removing bright SPIRE sources \citep{Nguyen10}.  Forty AzTEC sources are detected at 500 $\micron$ while 8 AzTEC sources do not have a significant counterpart in any of the SPIRE bands. The redshift tracks of a modified black-body with $T_{\rm dust} = 35$~K and typical starburst galaxies are also overlaid.  MM-J1545 is situated in a region of the plot where $z \approx 4$--5 galaxies are actually observed or are expected from redshift tracks of SED models.  Similarly, the 1.1-mm to 350-$\micron$ color of MM-J1541 is consistent with those found in $z \approx 3$ galaxies.

In the right panel of Figure~\ref{fig:ccd}, we show the $S_{\rm 350\mu m} / S_{\rm 250\mu m}$ and $S_{\rm 500\mu m} / S_{\rm 250\mu m}$ plots for MM-J1545 and MM-J1541, as well as the 40 AzTEC-selected sources in GOODS-South which are detected at 500 $\micron$.  Despite large uncertainties in flux density ratios, at least 16 of the GOODS-South sources (40\%) 
are consistent with 500-$\micron$ peakers that are detected at least at both 350 and 500~$\micron$, and up to 21 sources (53\%) 
may be 500-$\micron$ peakers if we include sources only detected at 500 $\micron$.  At least 3 sources meet the criterion, $S_{\rm 250\mu m} < S_{\rm 350\mu m} < S_{\rm 500\mu m}/1.3$, which is used to select $z \gtrsim 6$ candidates by \citet{Riechers13}.  The FIR-to-mm color of MM-J1545 is overall consistent with those of the 500-$\micron$ peakers but slightly bluer than the $z \gtrsim 6$ criterion, while the color of MM-J1541 is in good agreement with those of $z \sim 3$ sources, supporting the redshift estimates discussed in \S~\ref{sect:photoz}.

As demonstrated by these color--color plots, 1.1-mm selected sources that are very red in the (sub)mm will offer a unique opportunity to investigate how frequently massive starbursts are triggered in the $z > 4$ universe, which places constraints on galaxy formation models for massive dusty starbursts.  Future follow-up studies using the Atacama Large Millimeter/Submillimeter Array (ALMA) are needed to investigate this further.


\subsection{Is MM-J1545 a First Hydrostatic Core?} \label{sect:fc}

How to securely distinguish between SMGs and prestellar cores is always an issue in identifying SMGs behind Galactic molecular clouds, because flux densities and FIR-to-submm colors of the brightest SMGs looks similar to those found in the low-luminosity end of prestellar cores.  While MM-J1541 is well isolated from the Lupus-I molecular cloud ($A_V \simeq 1$) and thus likely an extragalactic source, MM-J1545 is closer to the molecular cloud ($A_V \simeq 2.4$, see also Figure~\ref{fig:azteconspire}) and is worthy of assessing the possibility that MM-J1545 is a Galactic source.   We hereafter assume a distance to the Lupus-I cloud of $D = 150$~pc \citep{Comeron08}, where a $1''$ size corresponds to the physical scale of $7.3 \times 10^{-4}$~pc or 150~AU.

The concordance scenario of low-mass star formation \citep[][for a review]{Shu87} begins with a collapse of a sub-pc-sized gravitationally-bound molecular core, a so-called prestellar core\footnote{%
A compact dense molecular cloud core which lacks a central (proto)star, whether it is gravitationally bound or unbound, is referred to as a starless core.  Thereafter we use the term, prestellar core, to distinguish the advanced protostellar objects, and for simplicity we neglect whether it is bound or not.
}.  Prestellar cores are initially optically-thin to the thermal dust emission and isothermally collapses, and thus it is represented as a cold \citep[$\simeq 10$~K, e.g.,][]{Marsh14} and moderately dense ($n$(H$_2$) $\sim 10^5$~cm$^{-3}$) isothermal system.  Most of them are found in regions where high H$_2$ column densities of $N({\rm H}_2) \gtrsim 10^{22}$~cm$^{-2}$ are observed \citep[e.g.][]{Rygl13}.  The prestellar phase is followed by protostellar phases \citep[a.k.a., Classes 0, I, II, and III,][]{Shu87,Andre93}, in which mid-IR and/or NIR emission powered by a central protostar is always visible.  
A hypothetical small adiabatic region that is extremely-dense ($n({\rm H}_2) \sim 10^{10}$--10$^{13}$~cm$^{-3}$) and compact ($\sim 1$--10$^2$ AU), a so-called first hydrostatic core \citep{Larson69}, may occur at the center of a prestellar core in its final stage just before protostar formation, although the first core is expected to be short-lived \citep[$\sim 10^3$ yr; e.g.,][]{Saigo11, Tomida13} and the observational nature is still controversial \citep[e.g.,][]{Enoch10, Chen10, Pineda11}.

MM-J1545 is a very unusual source if this would be associated with the Lupus-I cloud.  The stark constraint on 24-$\micron$ flux density ($< 0.3$~mJy), the absence of a bright compact NIR source \emph{at the position of} the SMA source or extended reflection nebulosity (as seen in edge-on Herbig-Halo objects like HH30), the low (effective) dust temperature of $T_{\rm dust} \approx 7.1$~K (\S~\ref{sect:mmj1545}), and the non-detection of $^{12}$CO outflows virtually rule out the protostellar Class-0 and advanced phases, suggesting that MM-J1545 might be in the prestellar phase and thus the mass should be dominated by dense molecular gas.  C$^{18}$O $J$ = 1--0 and 2--1 emission lines toward this source (i.e., $z = 0$) were, however, not detected with the SMA and the NRO 45~m telescope (Figure~\ref{fig:45obs}), which places a constraint on molecular gas mass of $M_{\rm LTE} < 0.005\, M_{\sun}$ (\S~\ref{sect:mmj1545}).  In contrast, the 1.1~mm flux density of 44~mJy yields $M{\rm (H_2)} \sim 0.1\, M_{\sun}$ if this is a starless core [$T_{\rm dust} = 7.1$~K, $\kappa_{\rm d} = 0.1\,(\lambda / 250\,\micron)^{-\beta}$ g$^{-1}$~cm$^2$ \citep{Hildebrand83} with $\beta = 1.4$, and the dust-to-gas mass ratio of 100 are assumed].  This discrepancy between gas and dust masses indicates that MM-J1545 is not a prestellar core.

Furthermore, the SMA continuum observations at 890~$\micron$ and 1.3~mm revealed a very compact source, compared to usual prestellar cores (\S~\ref{sect:individual}).  As shown in Figure~\ref{fig:uvplot}, its visibility distribution does not show any evidence for an extended structure like a dense gas envelope, which is at least $\sim$0.01~pc and typically $\sim 0.1$~pc in size, corresponding to an angular size of $\sim 100''$ \citep[e.g.][]{Onishi02}.  In Figure~\ref{fig:uvplot}, we show Fourier transform of 890-$\micron$ brightness distribution predicted for a critical Bonnor--Ebert sphere \citep{Ebert55, Bonnor56} as a realization of a starless core in the Lupus-I cloud.  The gas temperature and central H$_2$ density is assumed to be 7.1~K and $5 \times 10^7$~cm$^{-3}$, respectively.  The total gas mass of $0.04\,M_{\sun}$ is chosen so that the total 890-$\micron$ flux density matches the observed one.  The overall visibility slope and amplitude at higher spatial frequencies are, however, largely deviated from the prediction of the Bonnor--Ebert model.

A possible Galactic interpretation of MM-J1545 might be a first hydrostatic core \citep{Larson69, Masunaga98, Masunaga00} in its very early phase, because its compact nature and the very cold ($T \approx 10$~K) and low-luminosity ($L_{\rm bol} \sim 10^{-3} L_{\sun}$) SED are consistent with those expected in a mostly-naked first core \citep{Tomida10}.  In this case, C$^{18}$O emission line is optically thick and the non-detection of C$^{18}$O is explained if a significant fraction of gas mass, which is inferred from the dust SED ($\sim 0.1\,M_{\sun}$), might be stored behind the C$^{18}$O photosphere.  In Figure~\ref{fig:uvplot}, we show the Fourier components of a 890-$\micron$ brightness distribution predicted for a first hydrostatic core, which is computed by radiation hydrodynamic simulations \citep{Tomida10, Saigo11}.  The first hydrostatic core is produced by gravitational collapse of a $0.3\,M_{\sun}$ Bonnor--Ebert sphere, and the mass and inclination angle are chosen so that the overall visibility amplitudes match the observed data.   Consequently, the unresolved cusp-like structure appeared at $\sqrt{u^2 + v^2} > 100$~k$\lambda$ and the extended envelope predicted from the first core model are consistent with the actual SMA visibility amplitudes.

Note that the adjacent compact NIR source, J1545B, and the observed JVLA 6~cm flux density at the SMA position cannot be explained consistently with the other data even in the first hydrostatic core models.  We also note that first hydrostatic cores should be extremely rare because of the short lifetime;  only one first core out of $\sim$100--1000 dense starless cores is expected, if comparing the lifetime with the dynamical time scale of starless cores ($\sim$0.1--1 Myr).   We therefore conclude that the $z \simeq 4$--5 lensed SMG is the most likely and naturally explained by the SED and geometry/spatial extent of the multi-wavelength counterpart to MM-J1545.


\section{Conclusions}\label{sect:conclusions}

We report serendipitous detections of two 1.1-mm bright extragalactic sources, MM J154506.4$-$344318 (MM-J1545) and MM J154132.7$-$350320 (MM-J1541), in the AzTEC/ASTE survey of the Lupus-I star-forming region.  These sources are likely located at cosmological distances at $z \simeq 4$--5 (MM-J1545) and $z \simeq 3$ (MM-J1541).  
The main results are as follows:
\begin{enumerate}

\item MM-J1545 is the brightest ($S_\mathrm{1.1mm} = 43.9 \pm 5.6$~mJy) of $\approx 1400$ SMGs identified through the whole AzTEC 1.1-mm galaxy surveys.  SMA (890~$\micron$ and 1.3~mm) interferometry confirms the exact position, and photometry from VLA, ATCA, NMA, and \textit{Herschel}, in addition to AzTEC/ASTE and SMA, constrains the SED well, which is in good agreement with a single gray-body with $T_\mathrm{dust} / (1+z) = 7.1 \pm 0.3$~K and $\beta = 1.4 \pm 0.1$ (\S~\ref{sect:mmj1545}).  The SED fits to the photometry at $\ge 24~\micron$ indicate a redshift of $z \simeq 3.4$--5.6 (a combination of 68\% 
confidence intervals of SED fits using the Arp~220, average SMG, and SMM~J2135 templates) (\S~\ref{sect:photoz}).  This is also supported by the (sub)mm color analysis, in which we show that MM-J1545 is situated in a region where $z \simeq 4$--5 dusty galaxies are expected on the $S_\mathrm{500\mu m}/S_\mathrm{1.1mm}$--$S_\mathrm{350\mu m}/S_\mathrm{1.1mm}$ diagram (\S~\ref{sect:color}).  A faint NIR object, J1545B, is identified $0\farcs 9$ east of the SMA 890-$\micron$ peak, which is likely a foreground lensing object that amplifies MM-J1545.  Although it is difficult to constrain the SED of J1545B, the Einstein radius would be $\approx 0\farcs 9$ if the lens redshift is $z_\mathrm{L} \approx 0.5$.  Even if this is the case and  the magnification factor might be as high as 10, the demagnified FIR luminosity is still extreme ($L_\mathrm{FIR} \sim 10^{13}\, L_{\sun}$), suggesting that MM-J1545 is intrinsically a hyper-luminous galaxy at $z \simeq 4$--5 (\S~\ref{sect:gravlens}).

\item MM-J1541 is identified as the fourth brightest 1.1-mm source ($S_\mathrm{1.1mm} = 27.1 \pm 5.0$~mJy) of the whole AzTEC surveys. This object is also seen in the SPIRE 250--500~$\micron$ and MIPS 24~$\micron$ bands (\S~\ref{sect:mmj1541}).  The inferred photometric redshift ranges $z = 2.1$--4.1 (a combination of 68\% 
confidence intervals of SED fits using the Arp~220, M~82, average SMG, and SMM~J2135 templates) (\S~\ref{sect:photoz}).  Again, this is also supported by the AzTEC--SPIRE color analysis (\S~\ref{sect:color}).  An optical/NIR image (J1541B) is clearly offset from the 24-$\micron$ centroid by $\approx 3''$, and the SED of J1541B is consistent with a bright galaxy at $z_\mathrm{L} = 0.26^{+0.29}_{-0.13}$, suggesting that J1541B gravitationally magnifies the background SMG, MM-J1541.  Gravitational lens modeling using the Faber-Jackson relation and a singular isothermal sphere shows the magnification factor of MM-J1541 is moderate ($\mu_\mathrm{g} \approx 1.2$), suggesting that MM-J1541 might be an intrinsically hyper-luminous star-forming galaxy with $\log{(L_\mathrm{FIR}/L_{\sun})} \sim 13.7$--13.9 (\S~\ref{sect:gravlens}).

\item The brightest-end ($S_\mathrm{1.1mm} > 25$~mJy) of the 1.1-mm cumulative number counts is constrained by MM-J1545 and MM-J1541, in addition to another two sources from the literature;  $N(>\!25\,\mathrm{mJy}) =  0.70^{+0.56}_{-0.34}$~deg$^{-2}$, $N(>\!32.5\,\mathrm{mJy}) = 0.53^{+0.51}_{-0.29}$~deg$^{-2}$, and $N(>\!40\,\mathrm{mJy}) = 0.35^{+0.46}_{-0.23}$~deg$^{-2}$.  The slope at $S_\mathrm{1.1mm} \gtrsim 20$~mJy is shallower than the deeper part of the source counts obtained from general deep fields such as GOODS and COSMOS.  The amplitude and slope at the brightest-end is consistent not only with that properly scaled from the 1.4-mm counts obtained by the South Pole Telescope survey, which are thought to be dominated by strongly-lensed SMGs, but also with a model prediction where flux magnification due to strong gravitational lensing is accounted for.  This suggests that a substantial fraction of $S_\mathrm{1.1mm} > 25$~mJy sources may be gravitationally amplified.

\item  The overall SED from the optical to the radio and the spatial structure of MM-J1545 are explained neither by a local prestellar core nor a protostellar object, although it is found toward a relatively high H$_2$ column region of the local molecular cloud.  A possible explanation for a Galactic object might be a first hydrostatic core with no envelope structure.  Even in this case, however, neither the 6-cm continuum emission nor the NIR compact object, J1545B, $0\farcs 9$ away from MM-J1545 can be explained by any models for a first hydrostatic core.  On the other hand, a $z \simeq 4$--5 SMG strongly lensed by J1545B naturally accounts for all of the observed properties.  Hence we conclude that MM-J1545 is not a first hydrostatic core but a galaxy at a cosmological distance.  

\end{enumerate}

Unexpectedly, an extremely-bright SMG at $z > 3$ and a Galactic low-mass dense starless core (e.g., an exposed first hydrostatic core) could be similar in the mid-infrared to millimeter spectral energy distributions and spatial structures at least at $\gtrsim 1''$.  This indicates that it is necessary to distinguish the two possibilities by means of broad band photometry from the optical to centimeter, when a compact object is identified toward Galactic star-forming regions.   (Sub)millimeter spectroscopy and/or sub-arcsec imaging of the object will help to determine the redshift and the presence of gravitational magnification, which will be able to carried out using ALMA.


\acknowledgments

We are grateful to the ASTE team for making the observations possible.  We thank M.\ B\'{e}thermin and I.\ Shimizu for making the model number counts available.
The study is partially supported by KAKENHI (No.\ 25103503).
YS and KT are supported by JSPS Research Fellowship for Young Scientists (No.\ 09J05537 and 09J00159, respectively).
This work is based on observations and archival data made with the following telescopes and facilities: 
The ASTE telescope is operated by National Astronomical Observatory of Japan.
The Australia Telescope Compact Array is part of the Australia Telescope National Facility which is funded by the Commonwealth of Australia for operation as a National Facility managed by CSIRO.
\textit{Herschel} is an ESA space observatory with science instruments provided by European-led Principal Investigator consortia and with important participation from NASA. 
NRAO is a facility of the NSF operated under cooperative agreement by Associated Universities, Inc.
The 45-m radio telescope and the Nobeyama Millimeter Array are operated by Nobeyama Radio Observatory, a branch of National Astronomical Observatory of Japan.
The Submillimeter Array is a joint project between the Smithsonian Astrophysical Observatory and the Academia Sinica Institute of Astronomy and Astrophysics and is funded by the Smithsonian Institution and the Academia Sinica.
The \textit{Spitzer} Space Telescope is operated by the Jet Propulsion Laboratory, California Institute of Technology under a contract with NASA.
Subaru Telescope is operated by National Astronomical Observatory of Japan.
The United Kingdom Infrared Telescope is operated by the Joint Astronomy Centre on behalf of the Science and Technology Facilities Council of the U.K.
This publication makes use of data products from the Two Micron All Sky Survey, which is a joint project of the University of Massachusetts and the Infrared Processing and Analysis Center/California Institute of Technology, funded by the National Aeronautics and Space Administration and the National Science Foundation.

{\it Facilities:} \facility{ASTE (AzTEC)}, \facility{ATCA}, \facility{\textit{Herschel} (SPIRE)}, \facility{VLA}, \facility{NMA}, \facility{Nobeyama 45\,m (T100, BEARS)}, \facility{SMA}, \facility{\textit{Spitzer} (IRAC, MIPS)}, \facility{Subaru (MOIRCS)}, \facility{UKIRT (WFCAM)}, \facility{WISE}.

\end{document}